\documentclass[twocolumn]{aastex61}

\newcommand\aastex{AAS\TeX}

\submitjournal{ApJ}

\shorttitle{\aastex\ transit observations of PH$_3$ and H$_2$S}
\shortauthors{Wang et al.}

\begin{document}

\title{Modeling synthetic spectra for transiting extrasolar giant planets: detectability of H$_2$S and PH$_3$ with JWST}

\correspondingauthor{Dong Wang}
\email{dw459@cornell.edu}

\author{Dong Wang}
\affiliation{Department of Astronomy, Space Sciences Building, Cornell University,
14853, Ithaca, NY, USA}

\author{Yamila Miguel}
\affiliation{Observatoire de la C√¥te d‚ÄôAzur Boulevard de l‚ÄôObservatoire, CS 34229
06304 NICE Cedex 4, France}

\author{Jonathan Lunine}
\affiliation{Department of Astronomy, Space Sciences Building, Cornell University,
14853, Ithaca, NY, USA}



\begin{abstract}
 
JWST$'$s large aperture and wide wavelength coverage will enable it to collect the highest quality transit spectra observed so far. For exoplanetary atmospheres we expect to retrieve the abundance of the most abundant molecules, such as H$_2$O, CO, and CH$_4$. Other molecules, such as H$_2$S and PH$_3$, have been observed in Jupiter and Saturn but their chemistry and detectability in strongly irradiated planets is highly unknown. In this paper, we make the first effort to study their spectral features in solar composition atmospheres, and evaluate their detectability with JWST. We model the chemistry of phosphorus and sulfur in solar composition atmospheres. Our model includes the effect of vertical transport. Photochemistry effects are not included in our calculations. Using the abundance profiles, we model the JWST transmission and emission spectra for K=6.8 G-type star and for planets with cloud-free solar composition atmospheres. We find PH$_3$ is detectable at 3 sigma from transmission spectra of the simulated atmosphere with $T_{\rm eq}$ $<$ 500K using the NIRCam LW grism F444W mode with a total observing time of 28.8 hrs. H$_2$S is detectable at 3 sigma in the transmission and emission spectra for the simulated planet with $T_{\rm eq}$ $>$ 1500K using the NIRCam LW grism F322W2 mode with a total observing time of 24.0 hrs. Our results specifically highlight the importance of including H$_2$S for future abundances retrieval with JWST. The presence of clouds and hazes challenges the detections of PH$_3$ and H$_2$S, but H$_2$S features are still expected to be present in the emission spectra.

\end{abstract}

\keywords{planets and satellites, atmospheres --- 
planets and satellites, composition --- techniques, spectroscopic }



\section{Introduction} \label{sec:intro}
The atmospheres of exoplanets can be characterized by transit spectroscopy \citep[e.g.,][]{Seager00,Hubbard01}. The transit spectra contain information about the composition and vertical thermal structure of the atmospheres. Interpretation of the transit spectra has led to the discovery of atoms like sodium, potassium \citep{Charbonneau02,Redfield08,Jensen11,Sing12,Sing15,Nikolov14,Wilson15}, and molecules like H$_2$O \citep{Deming13,Huitson13,Mandell13,Crouzet14, McCullough14,Wakeford13,Wakeford17,Kreidberg14b,Kreidberg15,Evans16,Line16} in the atmospheres of hot Jupiters. Other molecules such as CH$_4$, CO, CO$_2$ were also reported in the literature to be detected, however, the detection of these molecules are not confirmed by later observations or other retrieval techniques \citep[e.g.,][]{Gibson11,Hansen14,Line14}. Currently, the determination of molecular abundances is limited by the quality of the transit spectra \citep{Burrow14}. 

James Webb Space Telescope (JWST)'s large aperture (6.5 m), wide wavelength coverage ($\lambda$ = 0.6 $\sim$ 28 $\micron$) and multiple instrument modes will ensure that it will collect the highest quality transit spectra \citep[e.g.,][]{Beichman14}. \citet{Greene16} simulated how well JWST observations can constrain the temperature-pressure profile and molecular abundances of H$_2$O, CH$_4$, CO, CO$_2$ and NH$_3$. Other molecules such as H$_2$S and PH$_3$ are not included in their calculations. However, H$_2$S and PH$_3$ are the primary carriers of sulfur and phosphorus in hydrogen-rich atmospheres \citep[e.g.,][]{Visscher06}, and they potentially contribute to the absorptions in the transit spectra. 

PH$_3$ has been observed in the atmospheres of Jupiter and Saturn \citep[e.g.][and references therein]{Fletcher09a}. The PH$_3$ observed in the upper troposphere and stratosphere are supplied by the vertical convection from deeper and hotter regions of the atmosphere where PH$_3$ is thermochemically stable. The same process may be at work in the exoplanets. H$_2$S was measured in the troposphere of Jupiter by the Galileo entry probe\citep{Irwin98,Wong04}. H$_2$S is the primary carrier of sulfur in the atmospheres of Jupiter and Saturn except above a few bars level where H$_2$S is removed by forming the NH$_4$SH cloud. For exoplanets with higher stellar irradiation, H$_2$S may not condense in the upper atmosphere. Therefore, H$_2$S can potentially contribute to the transit spectra of extrasolar giant planets.  

The non-equilibrium chemistry of phosphorus species was not explored in the context of exoplanets with hydrogen-rich atmospheres in the literature. The vertical mixing can drive the chemistry out of  equilibrium, just like the case in Jupiter and Saturn. Non-equilibrium chemistry of sulfur in extrasolar giant planets was studied by \citet{Zahnle09b}. From their calculations, H$_2$S is predicted to be the primary carrier of sulfur up to $\sim$ 10 mbar. Above 10 mbar, photochemistry is at work and the abundance of H$_2$S decreases at higher altitude.       

In this paper, we model the non-equilibrium chemistry of phosphorus in the hydrogen-rich atmospheres of exoplanets. We also model the non-equilibrium chemistry of C/N/O/S bearing species in order to get the vertical profiles of major molecules in the atmospheres. To evaluate whether H$_2$S and PH$_3$ can be detected by JWST transit observations, we modeled the synthetic transmission and emission spectra with simulated noise levels. 

The paper is organized as the follows. In section \ref{sec:method}, we describe our chemical model, synthetic spectra model, and JWST noise model. In section \ref{sec:abundances}, we present our results on the computed abundance profiles of major C/N/O/S/P bearing species. In section \ref{sec:noiseless_spectra}, we present the synthetic transit spectra for four planetary systems with different levels of stellar insolation. In section \ref{sec:noise_spectra}, we add simulated JWST noise into the synthetic spectra, and evaluate the detectability of H$_2$S and PH$_3$. In section \ref{sec:discussions}, we discuss the implications for JWST transit observations, and limitations of our model. In section \ref{sec:conclusions}, we present the conclusions of this paper. 

\section{Methodology}\label{sec:method}

In this section, we describe our methodology for modeling the synthetic JWST transit spectra. We first model the chemistry of C/N/O/S/P and identify major species in the atmospheres that are abundant and thus potentially important for the opacity. Then we model the noiseless primary and secondary transit spectra using the computed abundance profiles. Finally we model the transit spectra with simulated JWST noise, and determine whether certain molecules will be spectroscopically detectable by JWST. We detail our methodologies in what follows. 

\subsection{Chemical model}\label{subsec:chem_model}

We use a one-dimensional diffusion-kinetic model developed in \citet{Wang15,Wang16} to compute the vertical profiles of molecular abundances. The code solves the equation
\begin{equation}\label{eqn:DK}
\frac{\partial{Y_i}}{\partial{t}} = \frac{1}{\rho} \frac{\partial}{\partial{z}}(\rho K_{\rm eddy} \frac{\partial{Y_i}}{\partial{z}}) + P_i - L_i,
\end{equation} 
where $Y_i$ is the mass fraction of species $i$, $\rho$ is the density of the atmosphere, $z$ is the vertical coordinate, $K_{\rm eddy}$ is the vertical eddy diffusion coefficient, $P_i$ is the 
chemical production rate of species $i$, and $L_i$ is the chemical loss rate of species $i$. 
Two physical processes are modeled by the equation. One is the chemical production or loss of species $i$, and the other is the vertical transport of species $i$. The mixing ratio of species in the atmospheres is determined by the dynamic balance between these two physical processes. 

We neglect the effect of photochemistry. The effect on the chemical abundances is the photo-dissociation of hydrogen-bearing species (e.g., H$_2$O, CH$_4$, NH$_3$) and the production of photochemical products (e.g., C$_2$H$_6$, C$_2$H$_2$, HCN) \citep{Moses11,Moses13a,Venot12,Kopparapu12,Agundez14a,MK14}. Photochemistry changes the abundances only in the upper atmosphere that is at millibar levels. Therefore, we expect our computed abundance profiles are valid below $\sim$ 10 mbar. 

The diffusion-kinetic model requires three kinds of input. First is the temperature-pressure ($T-P$) profile; second is a list of thermodynamic properties and a list of reactions between these species; third is the elemental compositions and the eddy diffusion coefficient. We detail how we choose the inputs below. 
\begin{itemize}

\item $T-P$ profile: we compute the $T-P$ profile using the model developed in \citet{Parmentier14,Parmentier15}, which is a non-gray analytical model.   

\item Thermodynamic properties and reaction rates: the thermodynamic properties are used to compute the equilibrium abundances as well as the backward reaction rates. The thermodynamic properties are compiled from \citet{BR05}, \citet{McBride93}, \citet{DB00}, and \citet{Venot12thesis}. The kinetic network used for modeling the C/N/O/H chemistry is consisting of 108 species and 1000 reactions, originally from \citet{Venot12}. The H/P/O reaction network consists of 24 species and 175 reactions, originally from \citet{Twarowski95}. A more detailed description of the C/N/O/H and H/P/O reaction networks used in this paper can be found in \citet{Wang16}, and both reaction networks can be downloaded at the KIDA database (\url{http://kida.obs.u-bordeaux1.fr/networks.html}).  

\item Elemental abundances: we assume that the elemental composition of the atmosphere is solar. The solar elemental abundances are from \citet{Asplund09}. 

\item Eddy diffusion coefficient: $K_{\rm eddy}$ is used in the one-dimensional chemical models for parameterizing the vertical transport. There is no observational constraint on the eddy diffusion coefficient on exoplanets. However, its values can be approximated by multiplying the vertical convective velocity derived from 3-D General Circulation Models (GCM) with the pressure scale height \citep[e.g.,][]{Moses11,Venot12,Parmentier13}. This mixing length theory approximation has an uncertainty on the order of 10 in the estimated eddy diffusion coefficient \citep{Smith98}. In this paper, we choose to use a constant profile for the $K_{\rm eddy}$, with values equal to 1$\times$10$^9$ cm$^2$ s$^{-1}$ throughout the atmospheres. We explore the dependence on $K_{\rm eddy}$ in section \ref{sec:discussions}.
\end{itemize}

In each simulation, we provide the elemental abundances, the $K_{\rm eddy}$, and the $T-P$ profile to set up the code, then we initialize the $Y_i$ of species with chemical equilibrium mass fractions. $Y_i$ are evolved towards a steady state where the time derivative term in equation (\ref{eqn:DK}) is zero. In the code, we terminate the simulation once the relative changes of mole fractions in successive $\Delta$t is smaller than 1$\times$10$^{-3}$, where $\Delta$t is the overturning timescale, defined as $H^2/K_{\rm eddy}$, where $H$ is the pressure scale height at 1 bar level. The output is the vertical profiles of $Y_i$ for each species in the model.     

\subsection{Synthetic spectra model}
To simulate the synthetic spectra of transiting exoplanets, we modified the Smithsonian Astrophysical Observatory 1998 (SAO98) radiative transfer code (see \citet{ts76,tj02,kt09} and references therein for details). The line-by-line radiative transfer code calculates the atmospheric emergent spectra and also transmission of stellar radiation through the atmosphere with disk-averaged quantities at high spectral resolution.  The atmosphere is divided in different layers, where the transmission is calculated using Beer's law.  Updates include a new database with molecules relevant for giant planets that include H$_2$O, CH$_4$, CO, CO$_2$, NH$_3$, N$_2$, HCN, PH$_3$, H$_2$S taken from HITRAN \citep{r13} and HITEMP \citep{r10} database.  

The overall high-resolution spectrum is calculated with 0.1cm$^{-1}$ wavenumber steps. We smear them out to a resolving power of 100 to simulate the resolution that we will obtain with the MIRI instrument. For NIRISS and NIRCam, we assume the observed spectra are binned to a resolution of 100. The smoothing was done using a triangular smoothing kernel. 

We neglected the effect of cloud and hazes. Their influence on our results is discussed in the section \ref{sec:discussions}.

\subsection{JWST noise model}\label{subsec:noise}
\subsubsection{Greene et al. (2016) noise model}
The noise of primary and secondary transit spectra is simulated following the 
recipes in \citet{Greene16}. Here we provide a compact summary of the noise modeling methodology, along with parameters in the model, summarized in Table \ref{tab:noise_params}. The selected JWST observing modes are from Table 4 of \citet{Greene16}. We did not cover the NIRISS SOSS observing mode. The features of H$_2$S and PH$_3$ are not as strong as those of H$_2$O and hence we expect their effects to be masked by H$_2$O. Therefore, we do not expect this part of the spectrum to be relevant for detection of PH$_3$ and H$_2$S. The NIRCAM instrument with LW grism mode covers the wavelength 2.5 $\sim$ 5.0 $\micron$ with a native resolution of $\sim$ 1700; the MIRI instrument with slitless mode covers the wavelength 5.0 $\sim$ 11 $\micron$ with a resolution of $\sim$ 100. We adopted a cutoff at 11 $\micron$ for MIRI slitless mode because the transmission becomes low at longer wavelength \citep{Kendrew15}. The selected JWST modes provide a wavelength coverage between 2.5 and 11 $\micron$. We choose a binned resolution of R = 100 for all modes to ensure each bin contains enough photons in our simulation. 

There are four noise components: the signal photon shot noise, the background photon shot noise, the detector noise, and the systematic noise. The equations for computing each component are from \citet{Greene16}. For completeness, we present these equations below, and describe how we choose the parameter values in these equations.  
\begin{itemize}

\item The number of signal photons in each spectral bin is computed following the equation 
\begin{equation}\label{eqn:signal}
S_{\lambda} = F_{\lambda}A_{\rm tel}t\frac{\lambda^2}{hcR}\tau,
\end{equation}
where $S_{\lambda}$ is the number of signal photons in each spectral bin, $F_{\lambda}$ is the flux of the signal as received at the telescope, $A_{\rm tel}$ is area of the aperture of JWST, $t$ is the integration time, $R$ is the binned spectral resolution, and $\tau$ is the total system transmission. The integration time $t$ is adopted as the full transit duration $T_{14}$ (assuming equal integration time for both in and out transit). The transmission $\tau$ is sensitive to wavelength and is obtained from the JWST documentation (\url{https://jwst-docs.stsci.edu/display/JTI/NIRCam+Filters}). The signal flux $F_{\lambda}$ is measured at three configurations, namely, in-transit, out-transit, and in-eclipse. We assume the in-transit and out-transit integration time is unity. The signal shot noise is equal to the square root of $S_{\lambda}$.  

\item The background signal is computed following the equation
\begin{equation}\label{eqn:background}
B_{\lambda} = b_{\lambda}tA_{\rm pix}n_{\rm pix}R_{\rm native}/R, 
\end{equation}
where $B_{\lambda}$ is the background photon numbers in each spectral bin, $b_{\lambda}$ is the background electron flux, $A_{\rm pix}$ is the area subtended by each pixel, $n_{\rm pix}$ is two times the number of spacial pixels covered by the spectrum, and $R_{\rm native}$ is the native resolution of the spectrum before binning. The values of above parameters used in this simulation are summarized in Table \ref{tab:noise_params}. The background shot noise is equal to the square root of $B_{\lambda}$.  

\item The total detector noise in single transit observation is calculated as 
\begin{equation}
N_{\rm d,tot} = N_d\sqrt{n_{\rm pix}n_{\rm ints}R_{\rm native}/R},
\end{equation}
where $N_d$ is the total detector noise in one integration, and $n_{\rm ints}$ is the number of integrations in one transit observation. The parameter $n_{\rm ints}$ depends on the total transit duration, the brightness limit of each instrument mode, and the brightness of the star. The parameter values are summarized in Table \ref{tab:noise_params}.  

\item The systematic noise cannot be reduced by summing over more observations. We adopted the systematic noise floor as suggested by \citet{Greene16}, as presented in Table \ref{tab:noise_params}. 

\end{itemize}
The four noise components are combined quadratically to compute the total noise in each spectral bin for a single transit observation. 

\subsubsection{PandExo noise model}
PandExo (\url{http://pandexo.science.psu.edu:1111/}) is a tool for computing the error of a spectrum given the stellar SED, the planet spectrum, and the transit duration \citep{Batalha17}. It is built on top of the STScI's Pandeia exposure time calculator (ETC). PandExo does automatic optimization of groups and integrations. We use this tool to compute the errors on our simulated spectra, and as a validation of our parameters in the Greene et al. (2016) noise model. 

\section{Results for abundances profiles}\label{sec:abundances}

\begin{figure*}
\plotone{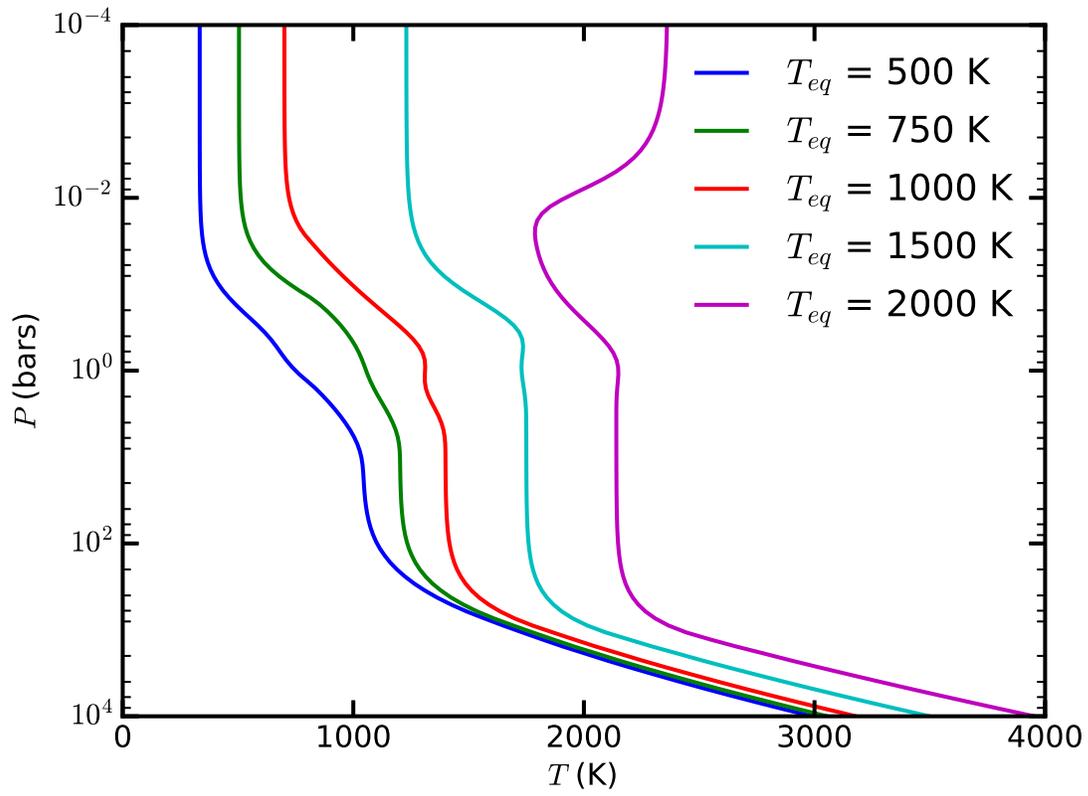}
\caption{Horizontally-averaged temperature-pressure profile for extrasolar giant planets computed using the approach in \citet{Parmentier14,Parmentier15}. Different lines correspond to different equilibrium temperatures, caused by the irradiation from the star.  
\label{fig:PT_profile}}
\end{figure*}

In this section, we present our results for the chemistry of C/N/O/S/P species. Temperature and pressure are the most important factors for determining the molecular abundances. The chemistry is very different for differently irradiated atmospheres. We simulate the atmospheres with different equilibrium temperatures (500 K, 750 K, 1000 K, 1500 K, 2000 K). The $T$-$P$ profiles used in the calculations are shown in Fig. \ref{fig:PT_profile}. The vertical eddy diffusion coefficient used is 1$\times$10$^9$ cm$^2$ s$^{-1}$, and the composition is assumed to be solar. In the following subsections, we present the computed vertical abundance profiles for $T_{\rm eq}$ = 500 K, 1000 K, 1500 K, and 2000 K. 
 
\subsection{Results for phosphorus species}\label{subsec:P_results}

\begin{figure*}
\gridline{\fig{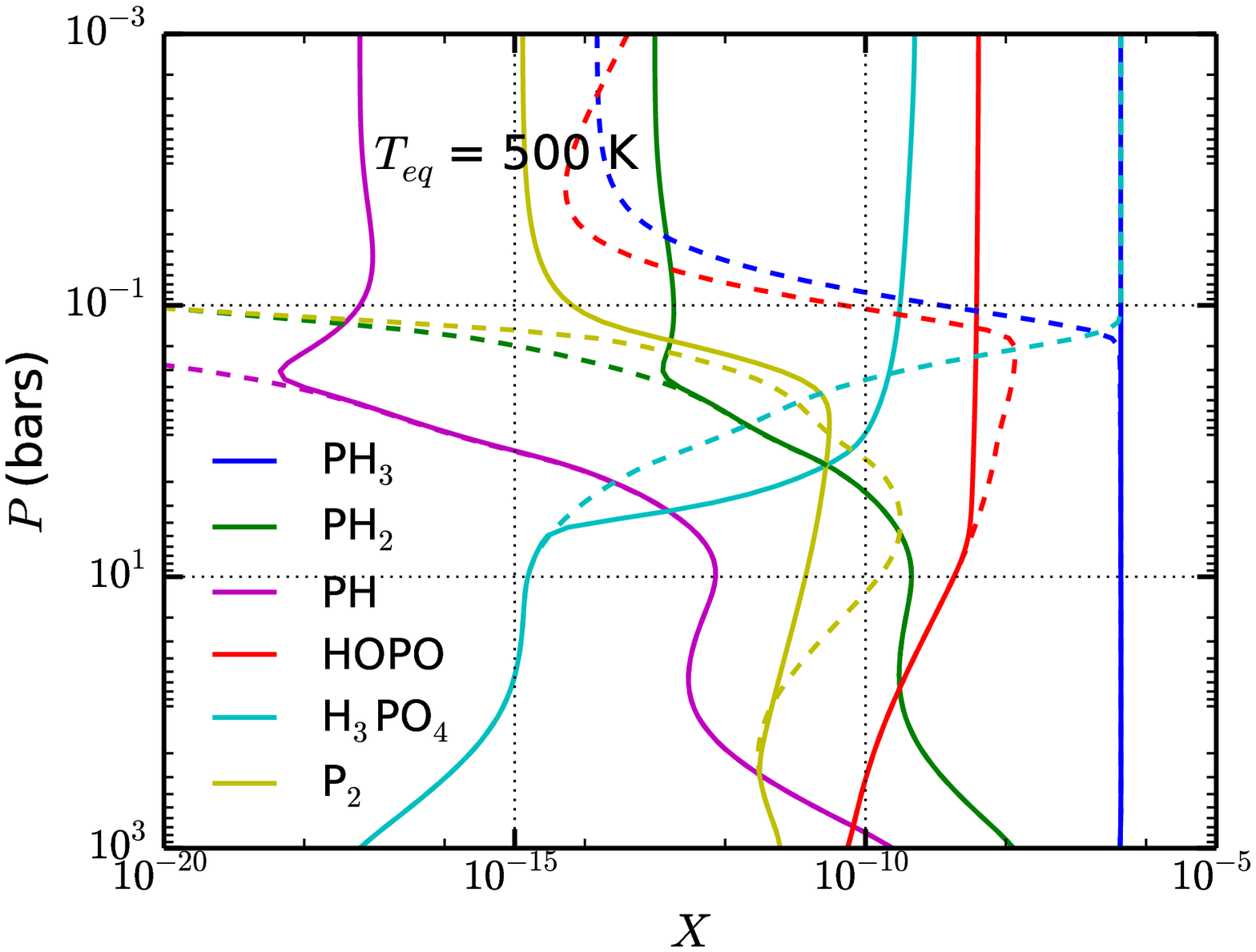}{0.5\textwidth}{(a)}
          \fig{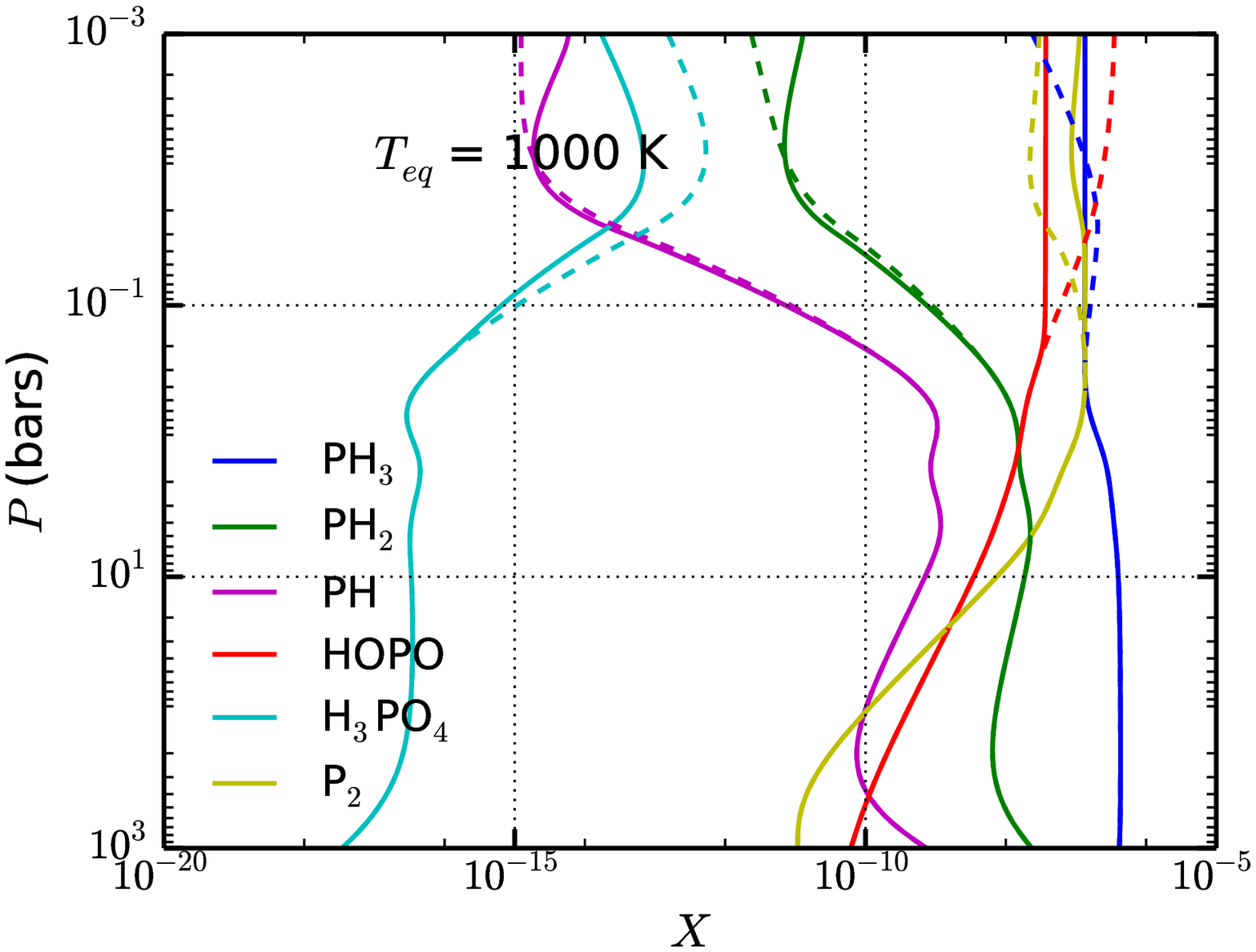}{0.5\textwidth}{(b)}
          }
\gridline{\fig{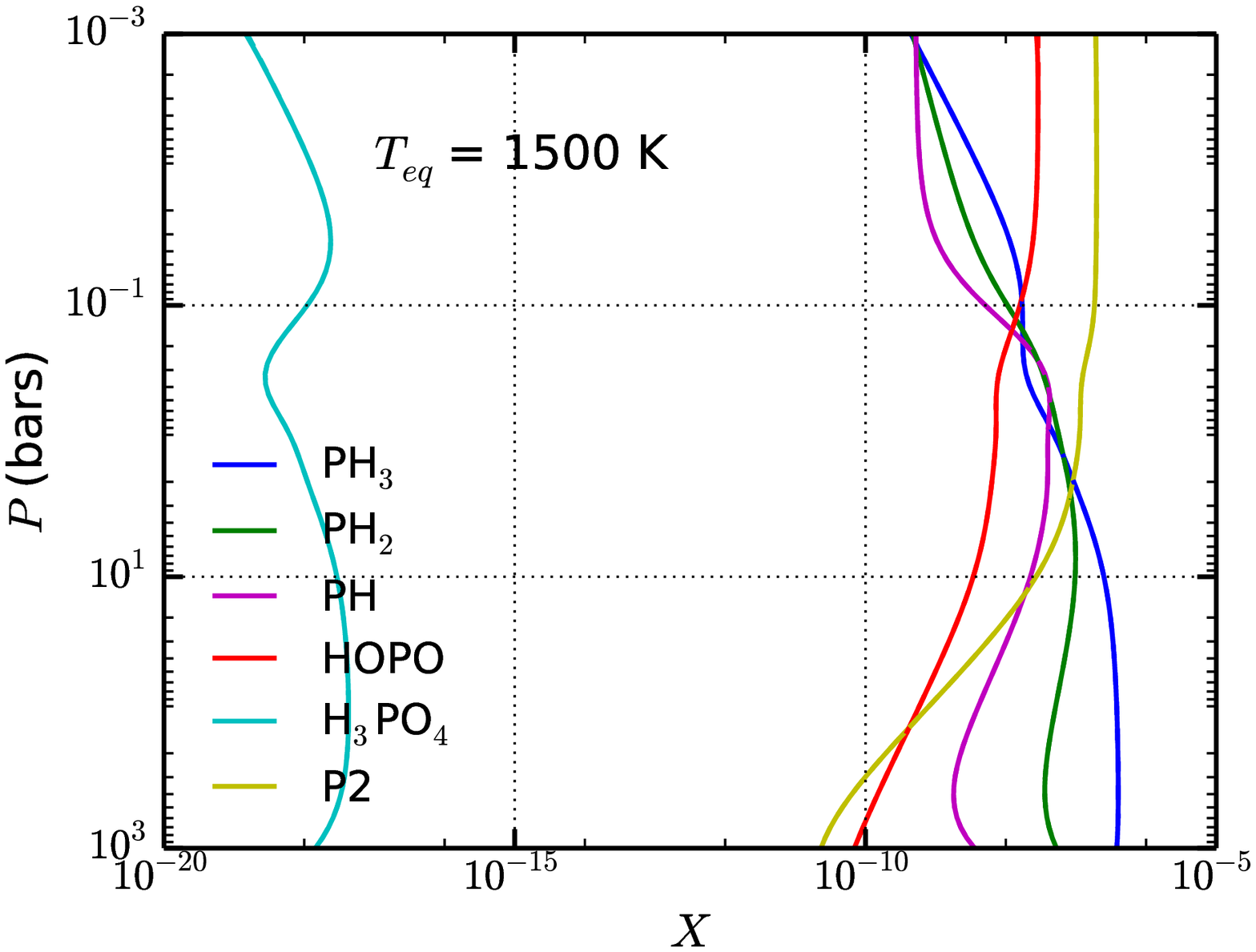}{0.5\textwidth}{(c)}
          \fig{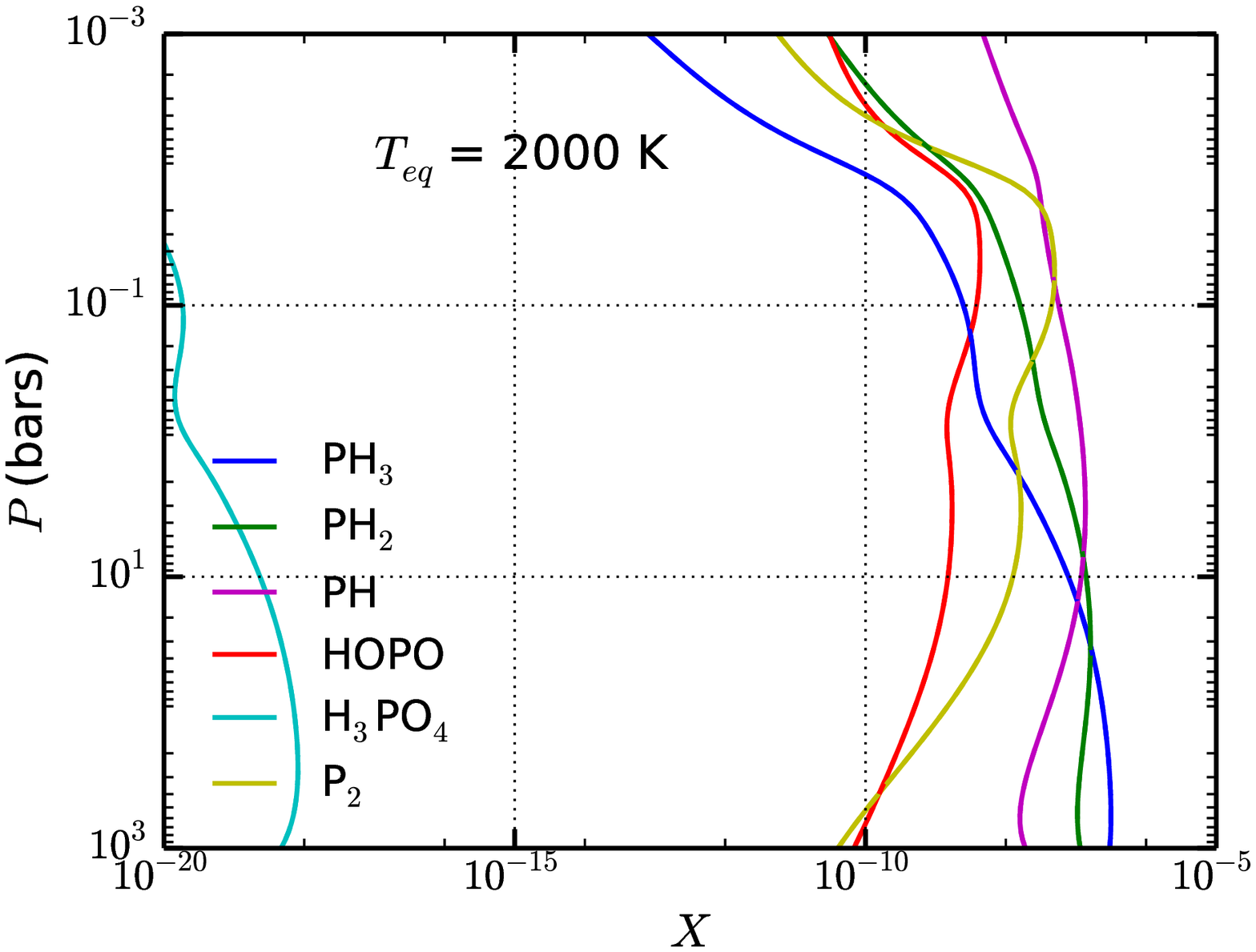}{0.5\textwidth}{(d)}
          }
\caption{Computed mole fractions of major phosphorus-bearing species (PH$_3$, PH$_2$, PH, HOPO, H$_3$PO$_4$, and P$_2$) in solar composition atmospheres. Solid lines show the abundances including the effect of vertical mixing, and the dashed lines show the abundances assuming local chemical equilibrium. The four plots correspond to different equilibrium temperatures: 
(a) $T_{\rm eq}$ = 500 K, (b) $T_{\rm eq}$ = 1000 K, (c) $T_{\rm eq}$ = 1500 K, (d) $T_{\rm eq}$ = 2000 K. The elemental abundance is assumed to be one solar. The eddy diffusion coefficient $K_{\rm eddy}$ is set at $1\times10^{9}$ cm$^2$s$^{-1}$. Due to the lack of photodissociation modeling, the profiles are valid below $\sim$ 0.1 bar.
\label{fig:phosphorus_chemistry}}
\end{figure*}

Assuming solar elemental abundances for phosphorus, hydrogen, and oxygen, we compute the abundance profiles of H/P/O bearing species for different levels of insolation. We present our results for the phosphorus chemistry in Fig. \ref{fig:phosphorus_chemistry}. The most abundant H/P/O bearing species are PH$_3$, PH$_2$, PH, HOPO, H$_3$PO$_4$, and P$_2$. For solar composition atmospheres with $T_{\rm eq}$ = 500 K and $T_{\rm eq}$ = 1000 K, the abundances are out of chemical equilibrium due to the effect of vertical mixing. For solar composition atmospheres with $T_{\rm eq}$ = 1500 K and $T_{\rm eq}$ = 2000 K, the abundances are in chemical equilibrium. The vertical mixing still exists, however, the mixing time scale is longer than the chemical timescale, and the abundances quickly re-equilibrate after mixing.  
The major phosphorus species are different for different levels of insolation. For a solar composition atmosphere with $T_{\rm eq}$ = 500 K, the dominant phosphorus-containing species is PH$_3$. This is similar to Jupiter, which has an equilibrium temperature at approximately 160 K. For a solar composition atmosphere with $T_{\rm eq}$ = 1000 K, PH$_3$ and P$_2$ are the most abundant phosphorus containing species above 1 bar. Below 1 bar, PH$_3$ is still the dominant species. At this temperature, part of the PH$_3$ is thermally decomposed into PH$_2$. The reactions between radicals can produce molecules with two or more phosphorus atoms such as P$_2$. At $T_{\rm eq}$ = 1500 K and 2000 K, the temperature is high enough that most PH$_3$ is thermally decomposed into PH$_2$ and PH. The photodissociation of PH$_3$ is potentially important at low pressure levels. For Jupiter and Saturn, the PH$_3$ abundance decreases near 0.1 bar level\citep{Fletcher09a}. There is no PH$_3$ photochemical modeling for hot Jupiters. In our case of solar composition Jupiter-size planets, we assume our abundance profile is valid below 0.1 bar. Above 0.1 bar, the profile is subject to change due to photodissociation. 

\subsection{Results for sulfur species}\label{subsec:S_results}

\begin{figure*}
\gridline{\fig{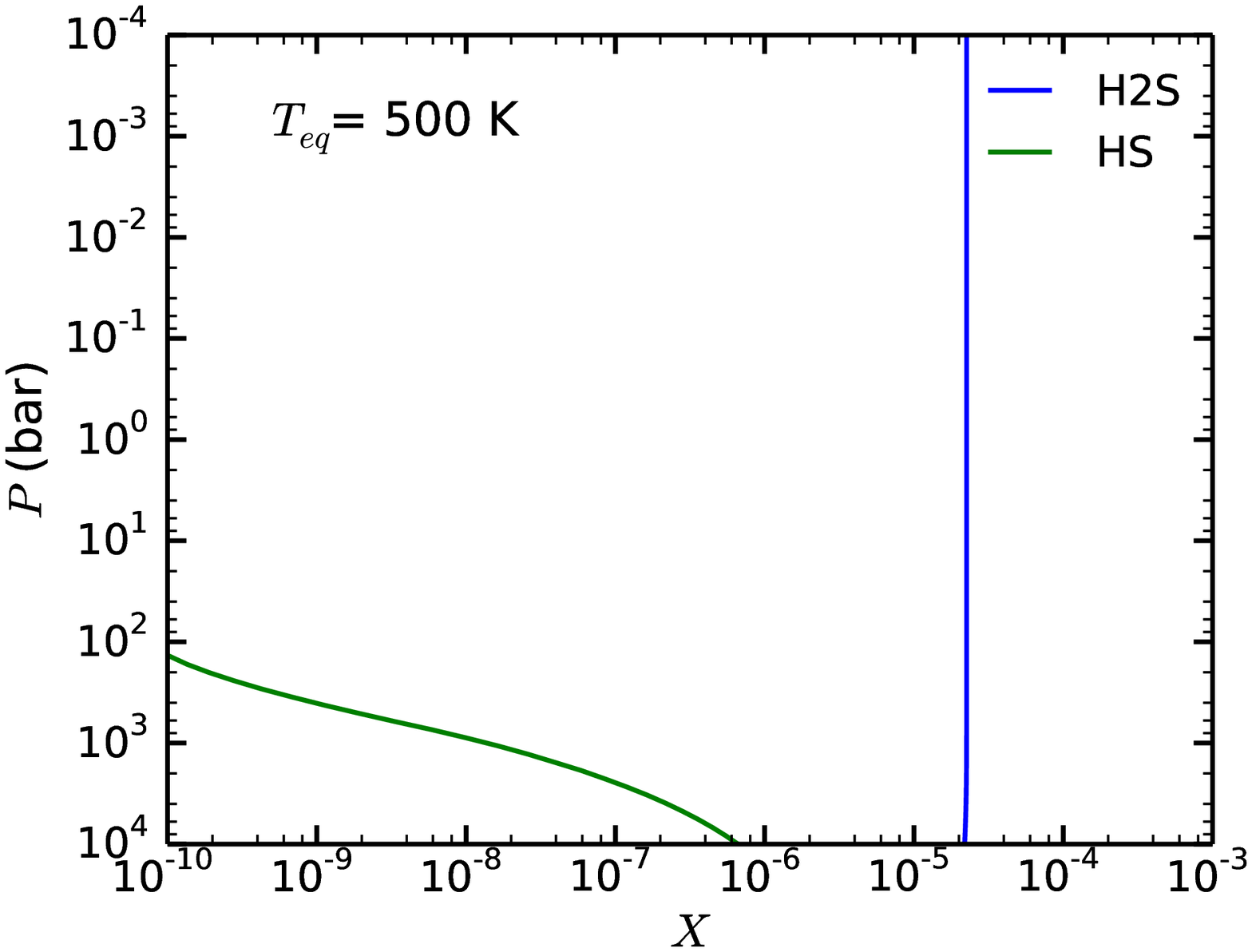}{0.5\textwidth}{(a)}
          \fig{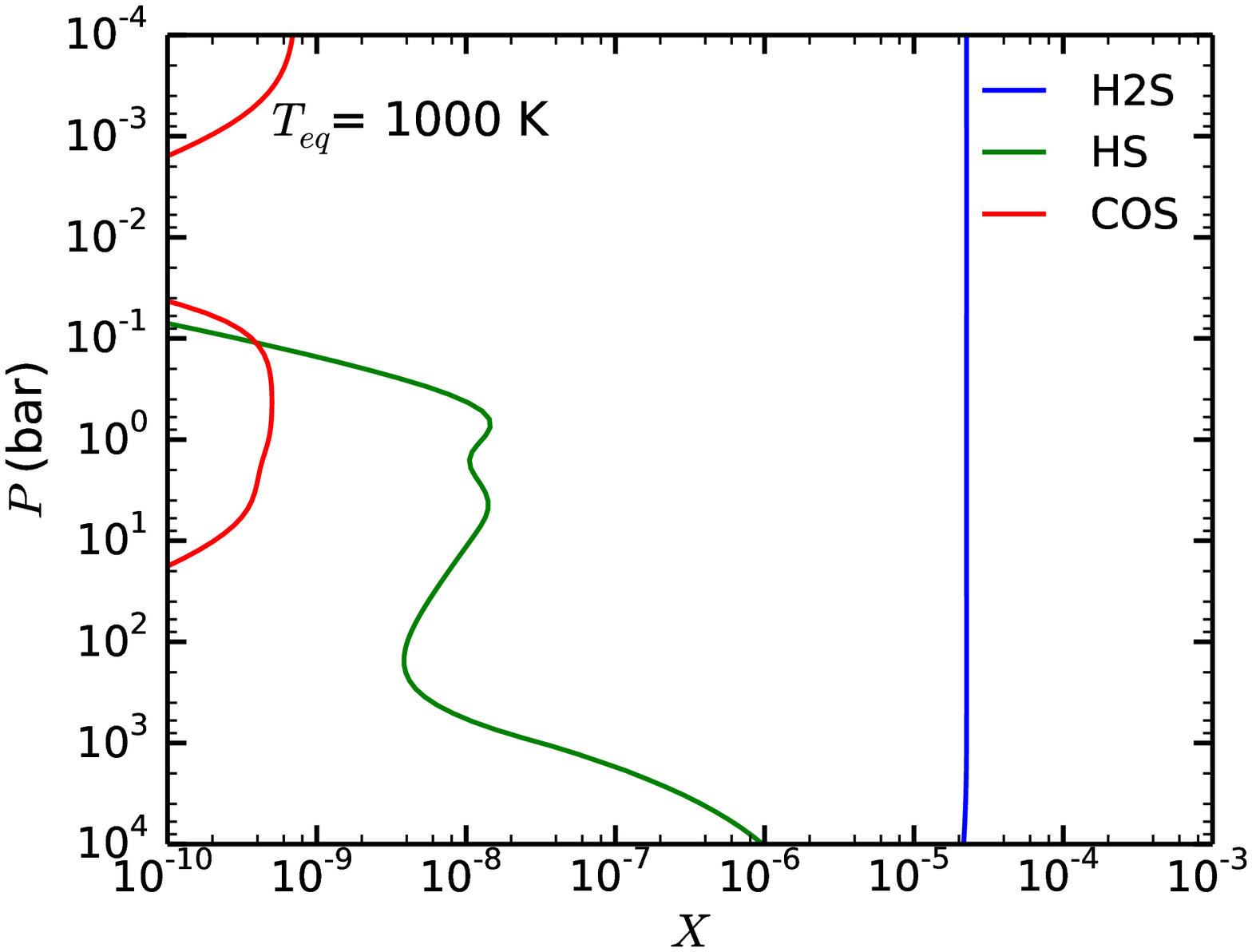}{0.5\textwidth}{(b)}
          }
\gridline{\fig{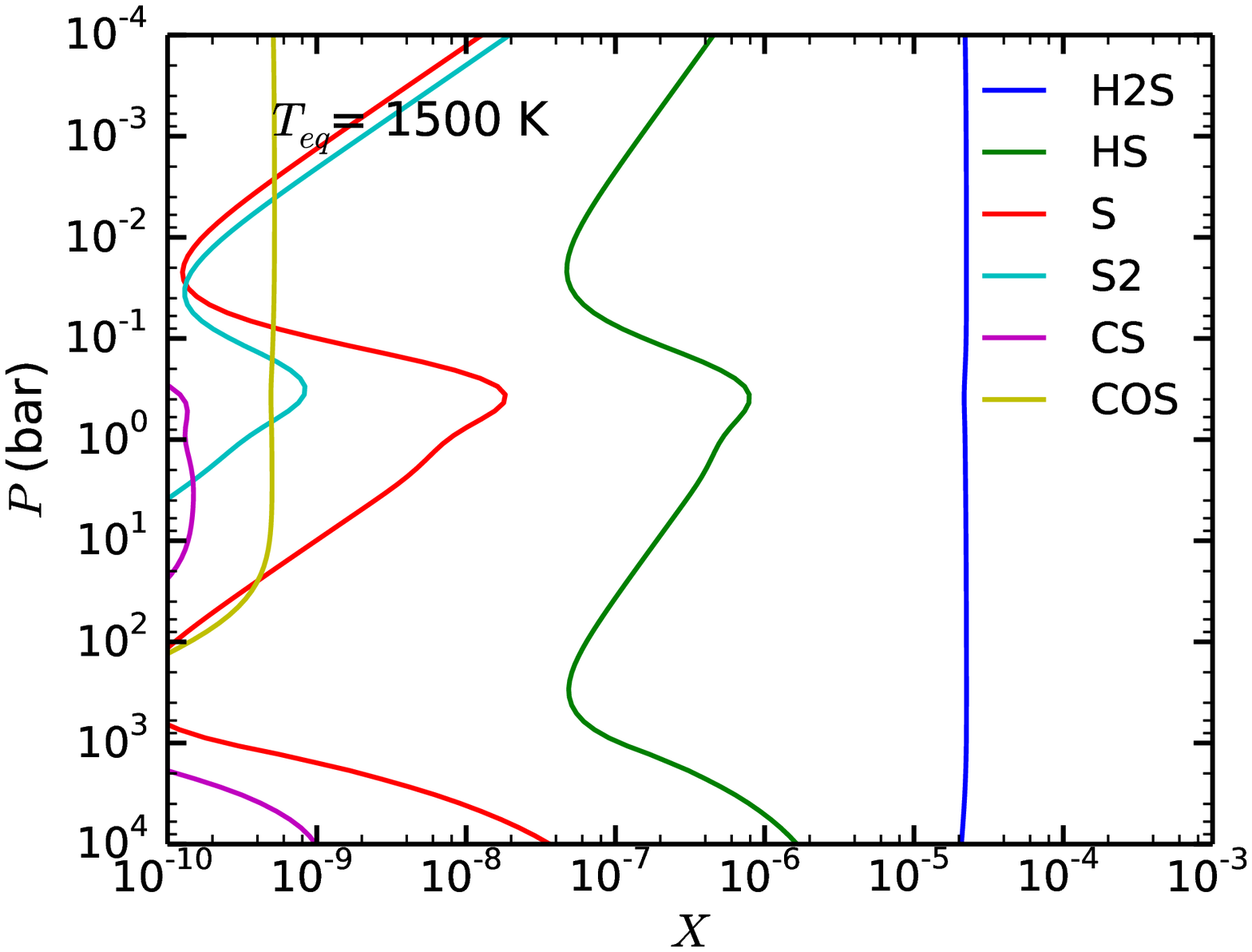}{0.5\textwidth}{(c)}
          \fig{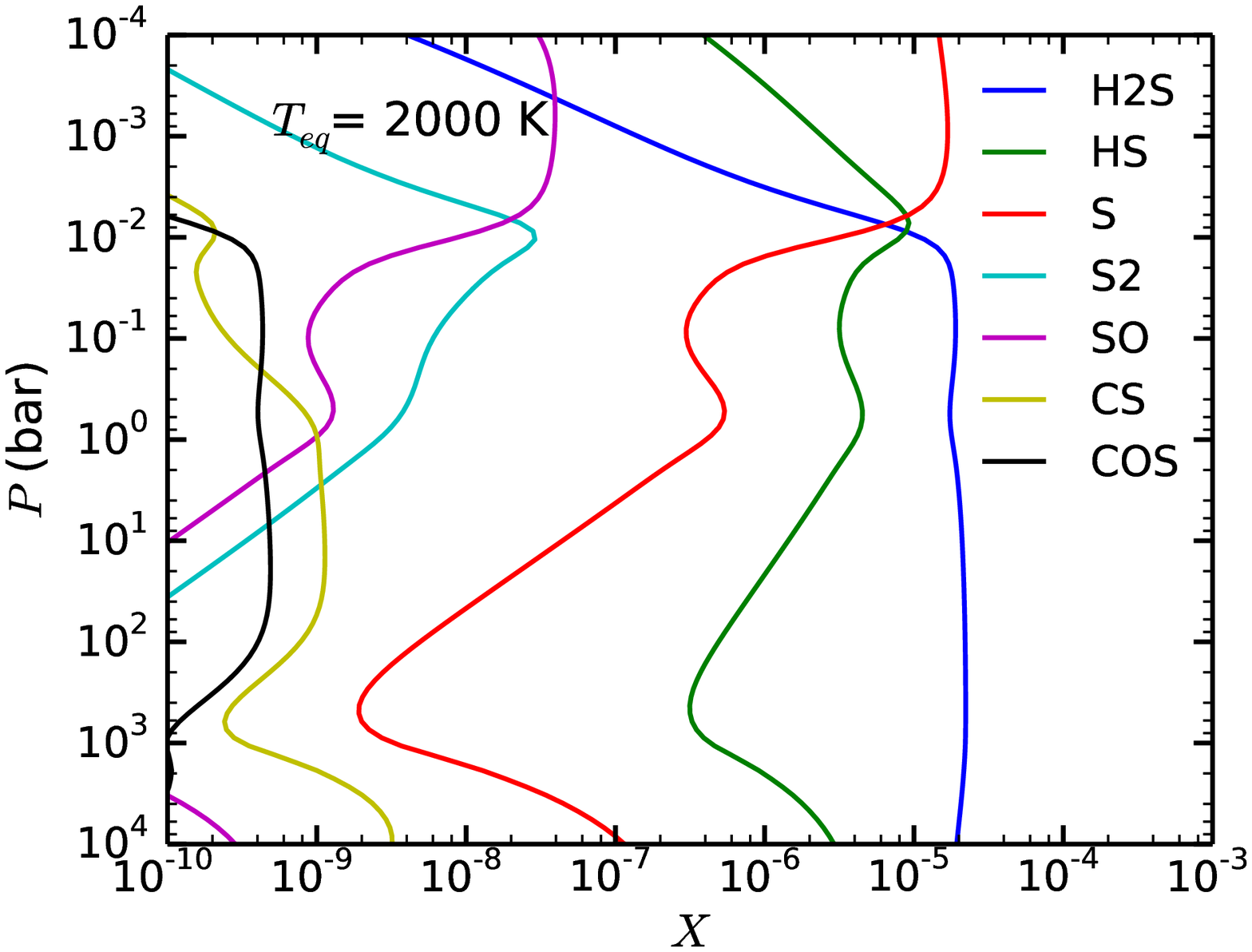}{0.5\textwidth}{(d)}
          }
\caption{Computed equilibrium mole fractions of major sulfur-bearing species in solar composition atmospheres. The four plots correspond to different equilibrium temperatures: 
(a) $T_{\rm eq} = 500 K$, (b) $T_{\rm eq} = 1000 K$, (c) $T_{\rm eq} = 1500 K$, (d) $T_{\rm eq} = 2000 K$.  The elemental abundance used here is one solar. The plot neglects the effects of photodissociation of H$_2$S, which may decrease the abundance of H$_2$S at $p$ $<$ 0.01 bar.
\label{fig:sulfur_chemistry}}
\end{figure*}

Assuming solar composition atmospheres, we compute the equilibrium abundances of sulfur species along the $T$-$P$ profile in order to identify the most abundant sulfur bearing species. We consider species H$_2$S, HS, H$_2$S$_2$, CH$_3$SH, S, S$_2$, SO, SO$_2$, CS, CS$_2$, COS, and SN. The results are shown in Fig. \ref{fig:sulfur_chemistry}. For solar composition atmospheres with $T_{\rm eq} = 500$ K, $1000$ K, and $1500$ K, H$_2$S is the dominant species at pressure levels between 1$\times$10$^{-4}$ bar and 1$\times$10$^4$ bar. For solar composition atmospheres with $T_{\rm eq} = 2000$ K, H$_2$S is the dominant species below 0.01 bar. Above 0.01 bar, atomic S is the dominant species. Since vertical mixing has the effect of homogenizing the abundances, the addition of vertical mixing into the model is not expected to change the result for H$_2$S. However, we ignore the effect of photochemistry, which may affect the vertical profile at low pressure levels. \citet{Zahnle09b,Zahnle16} have done photochemical modeling of sulfur species in the atmospheres of hot Jupiters and warm Jupiters. From their calculations, H$_2$S is largely photo-dissociated at $P \gtrsim 0.01$ bar, but remains the dominant sulfur carrier at $P \lesssim 0.01$ bar. \citet{Zahnle09b} showed that the abundances are not sensitive to temperature and insolation over the parameter ranges (T = 1200 $\sim$ 2000 K and I = 1 $\sim$ 1000). Following those results, we assume that H$_2$S is the dominant sulfur bearing species below 0.01 bar in our calculations. 

\subsection{Results for C/N/O/H species}

\begin{figure*}
\gridline{\fig{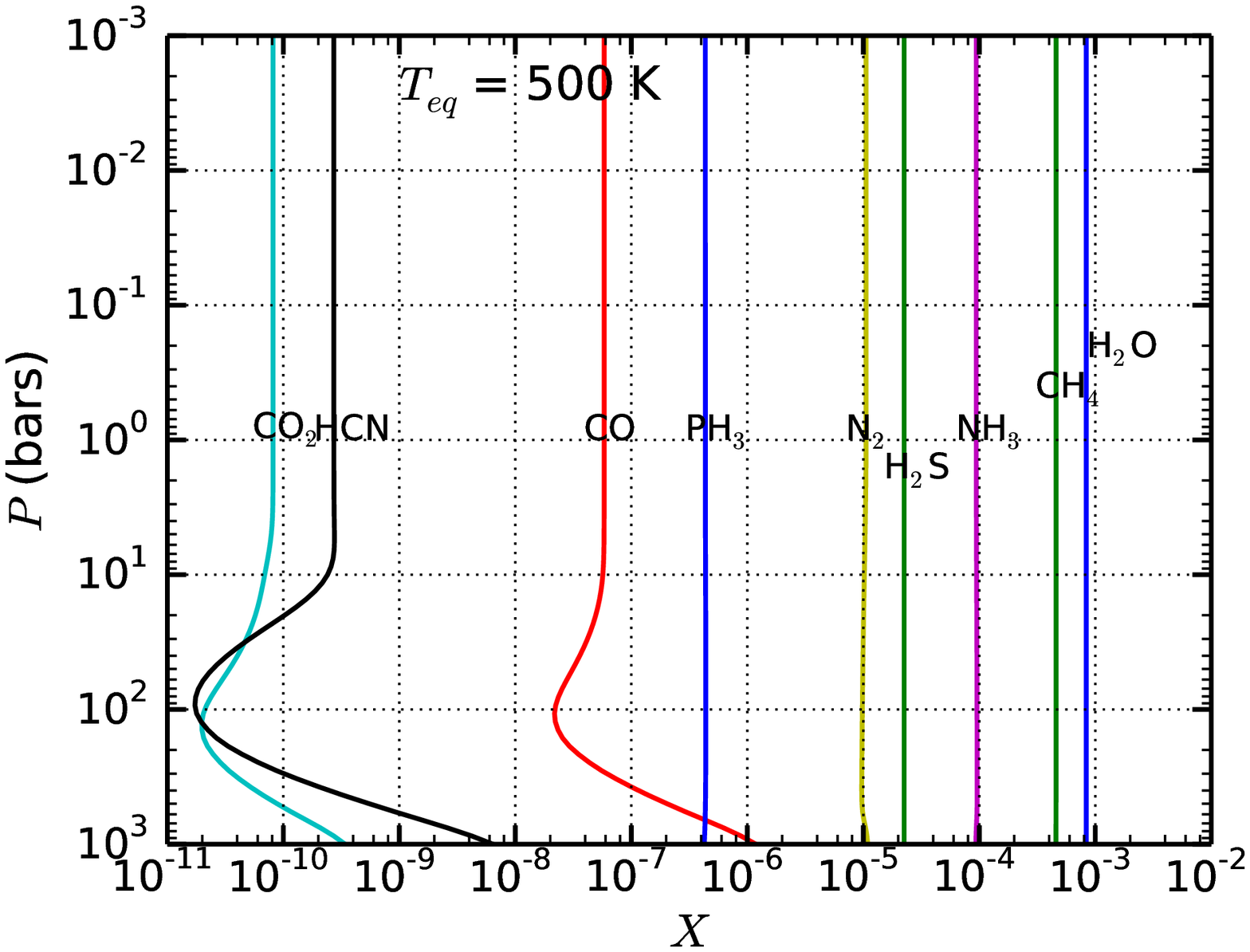}{0.5\textwidth}{(a)}
          \fig{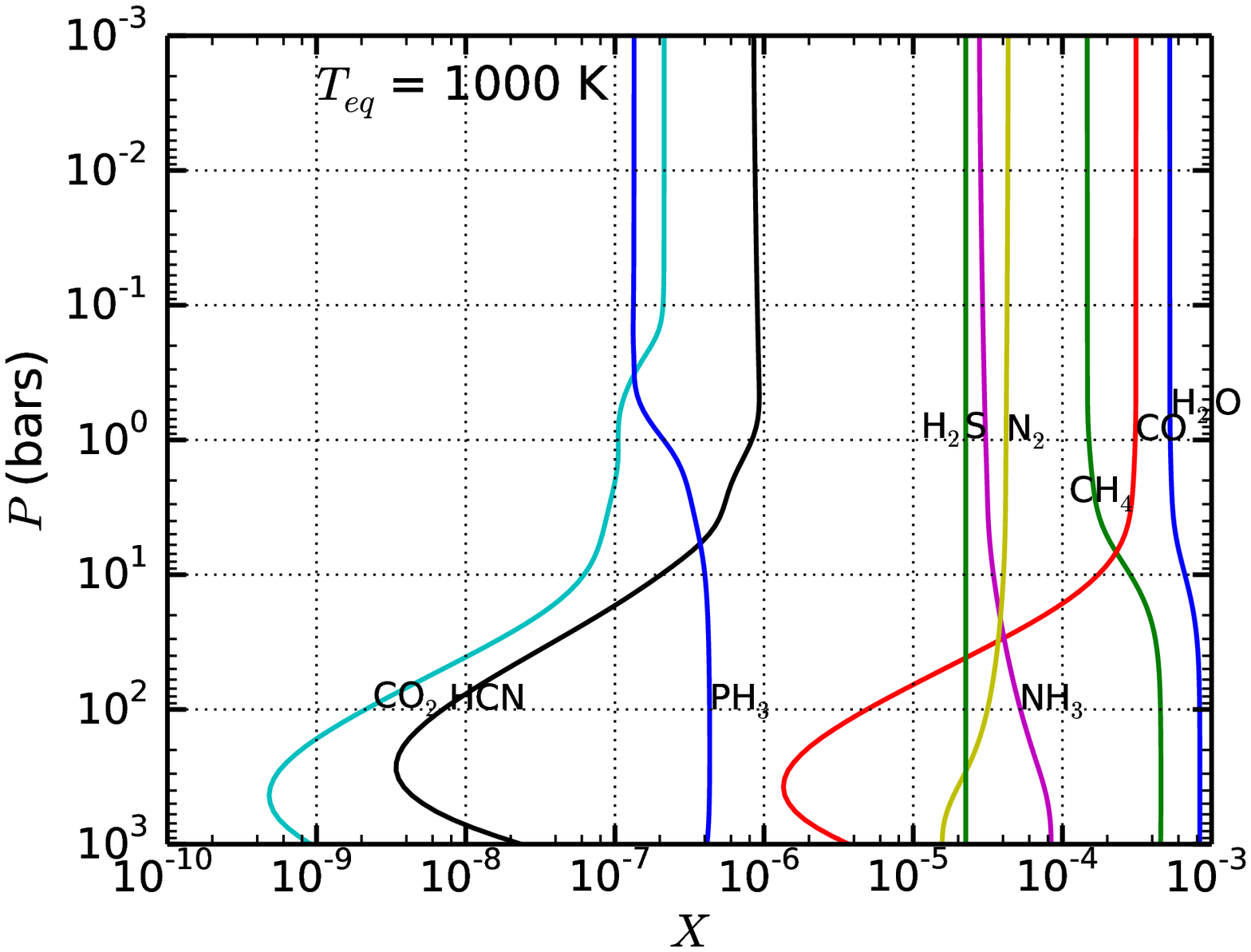}{0.5\textwidth}{(b)}
          }
\gridline{\fig{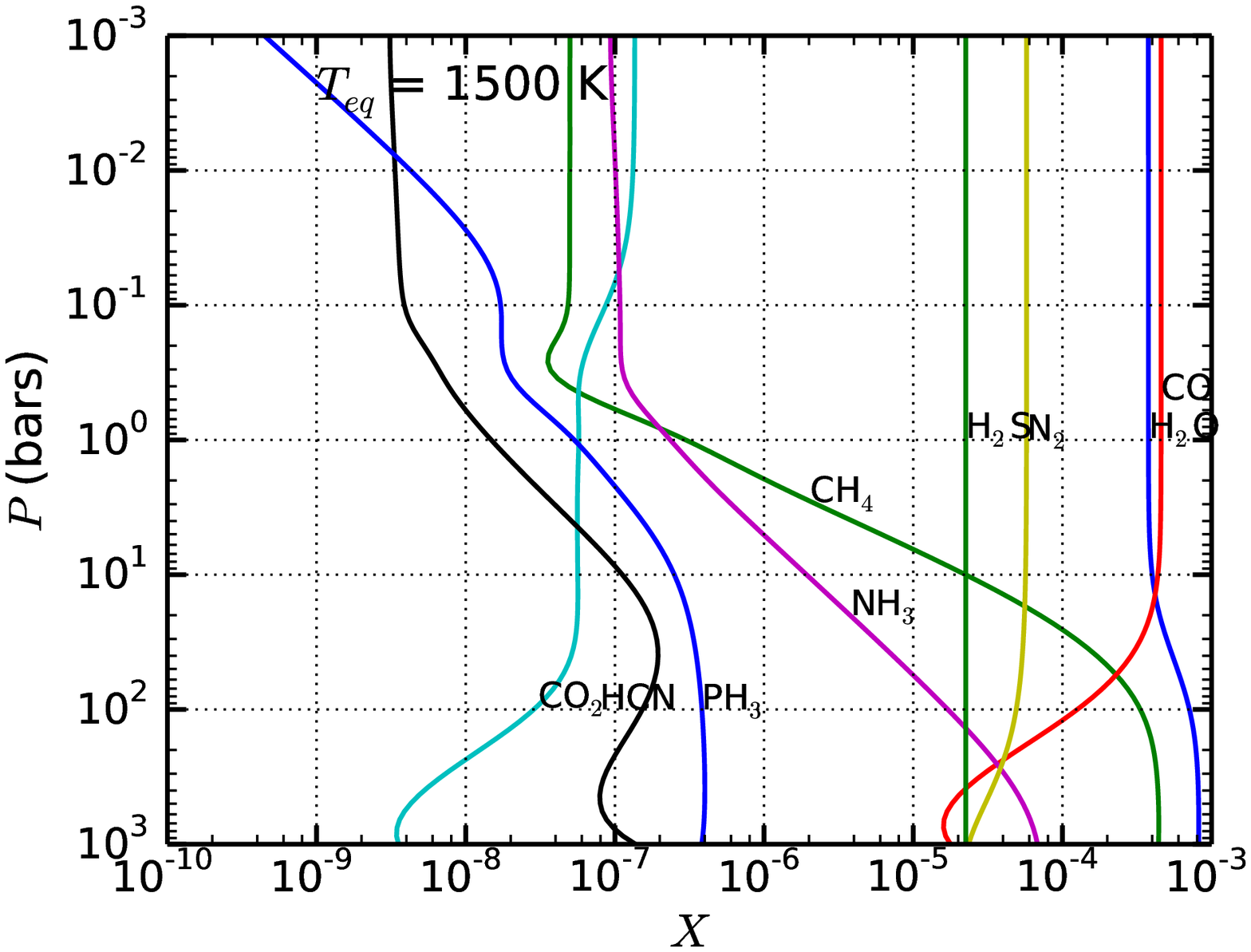}{0.5\textwidth}{(c)}
          \fig{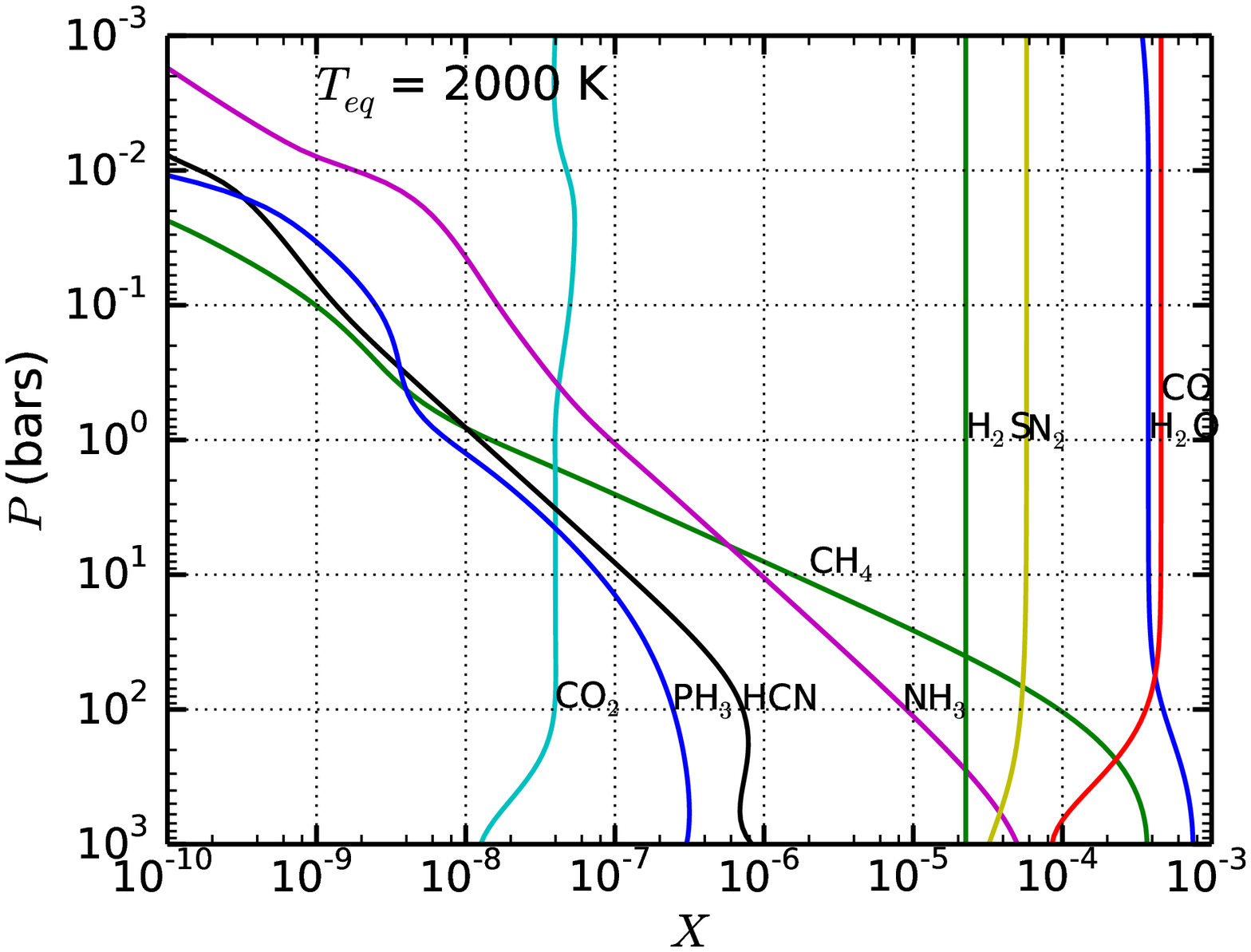}{0.5\textwidth}{(d)}
          }
\caption{Computed mole fractions of major C/N/O/S/P bearing species in solar composition atmospheres including the effect of vertical mixing. The four plots correspond to different equilibrium temperatures: 
(a) $T_{\rm eq}$ = 500 K, (b) $T_{\rm eq}$ = 1000 K, (c) $T_{\rm eq}$ = 1500 K, (d) $T_{\rm eq}$ = 2000 K. The elemental abundance used here is one solar. The vertical eddy diffusion coefficient used here is $K_{\rm eddy}$ = $1 \times 10^9$ cm$^2$ s$^{-1}$. The photodissociations of H$_2$O, CH$_4$ or NH$_3$ are expected to affect the profiles above $p$ $\sim$ 0.01 bar level, which is not included in this plot.  
\label{fig:major_species}}
\end{figure*}

It is necessary to carefully model the contribution of C/N/O/H bearing species to the transit spectra if we want to identify spectral features of H$_2$S and PH$_3$. Molecules such as H$_2$O or CO are more abundant than H$_2$S and PH$_3$, and thus contribute the most to the transit spectra. In order to find molecules that are more abundant than H$_2$S and PH$_3$, we performed independent calculations for C/N/O/H chemistry.  
Our results are in general consistency with results reported in the literature \citep[e.g.,][]{Moses11,Venot12,MK14,HS14}. We find the major C/N/O/H bearing molecules are H$_2$O, CO, CH$_4$, CO$_2$, N$_2$, NH$_3$, and HCN. These molecules must be included in the spectra calculation in order to cover all important opacity sources. We present our computed vertical profile of these molecules along with PH$_3$ and H$_2$S in Fig. \ref{fig:major_species}. The results are shown for four different levels of stellar insolation. 

For solar composition atmospheres with $T_{\rm eq} = 500$ K, the most abundant species are H$_2$O, CH$_4$, NH$_3$, H$_2$S, N$_2$ and PH$_3$. The abundances are nearly homogeneous in the vertical direction down to $\sim$ 10 bars. The atmosphere is strongly homogenized by vertical mixing, and species are in a disequilibrium state. CH$_4$ is the primary carbon-bearing species, H$_2$O is the primary oxygen-bearing species, NH$_3$ is the primary nitrogen-bearing species, H$_2$S is the primary sulfur-bearing species, and PH$_3$ is the primary phosphorus-bearing species. 

For solar composition atmospheres with $T_{\rm eq} = 1000$ K, the most abundant species are H$_2$O, CO, CH$_4$, N$_2$, NH$_3$, H$_2$S, HCN, and PH$_3$. Abundances are nearly homogeneous down to 1-10 bars due to the effect of vertical mixing. CO carries about 2/3 of the total carbon abundance, and CH$_4$ carries the other 1/3 of the total carbon abundance. H$_2$O is the dominant oxygen bearing species. N$_2$ and NH$_3$ each carries about 1/2 of the total nitrogen abundance. This temperature marks the transition temperature for CO/CH$_4$ conversion and N$_2$/NH$_3$ conversion. For $T_{\rm eq}$ $\lesssim$ 1000 K, CH$_4$ and NH$_3$ are the major carbon and nitrogen carriers; for $T_{\rm eq}$ $\gtrsim$ 1000 K, CO and N$_2$ are the major carbon and nitrogen carriers.   

For solar composition atmospheres with $T_{\rm eq} = 1500$ K, the most abundant species are CO, H$_2$O, N$_2$, and H$_2$S. CO is the primary carrier of both oxygen and carbon. The rest of the oxygen is in the form of H$_2$O. N$_2$ is the primary carrier of nitrogen. CH$_4$ and NH$_3$ are much less abundant in the atmospheres. 

For solar composition atmospheres with $T_{\rm eq}$ = 2000 K, the most abundant species are CO, H$_2$O, N$_2$, and H$_2$S. The abundances are nearly in chemical equilibrium due to the high temperature. CH$_4$, NH$_3$, and PH$_3$ are much less abundant. 

\subsection{Influence of insolation}\label{subsec:insolation}

\begin{figure*}
\plotone{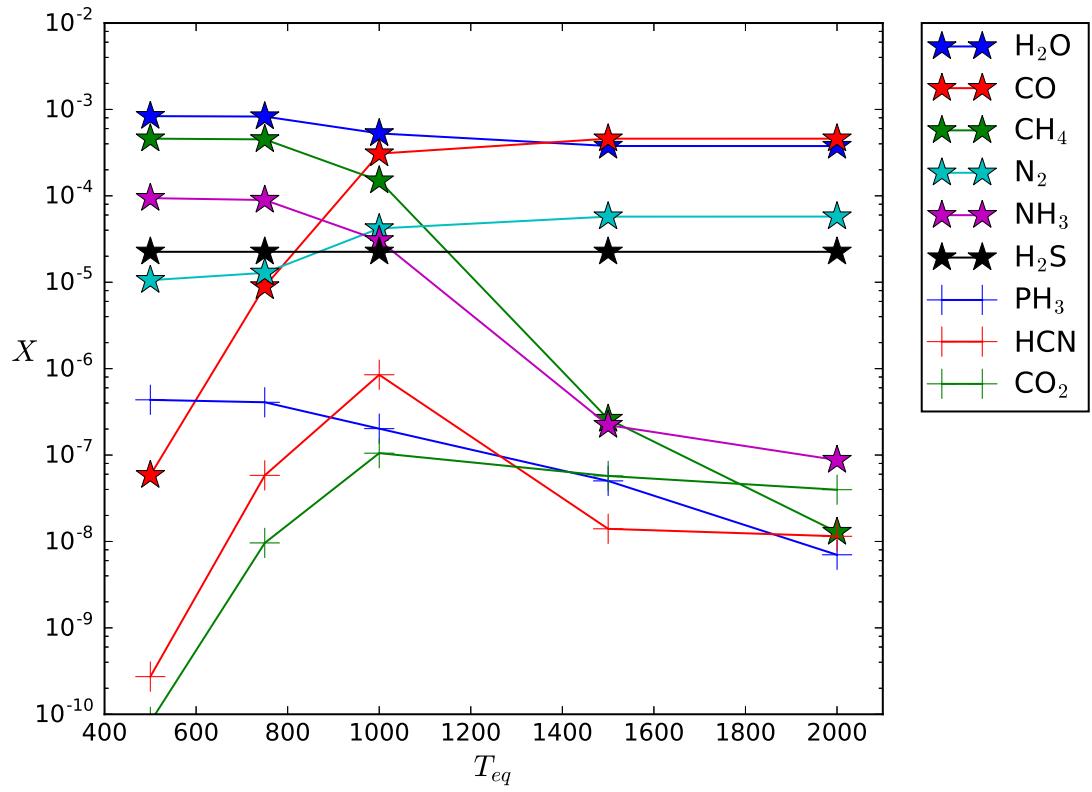}
\caption{Mole fractions at 1 bar as a function of the equilibrium temperatures of atmospheres. The species plotted are H$_2$O, CO, CH$_4$, CO$_2$, NH$_3$, N$_2$, HCN, H$_2$S, and PH$_3$. 
\label{fig:abundance_T}}
\end{figure*}

There are three regimes for the abundance profiles depending on the level of insolation. For highly irradiated atmospheres (e.g., $T_{\rm eq}$ $>$ 1500 K), the chemical abundances are in local chemical equilibrium. Therefore, when doing retrieval on atmosphere composition and $T$-$P$ profile, assumptions of chemical equilibrium should be valid. For moderately irradiated atmospheres (e.g., $T_{\rm eq}$ $<$ 1000 K), the vertical mixing tends to produce a homogeneous abundances in the atmospheres. It should be valid to assume a constant mixing ratio profile when doing atmospheric retrieval. In between is the transition regime when both chemical conversions and vertical mixing are important in the atmospheres. In this regime, the abundance profiles will depend on the vertical eddy diffusion coefficient as well as the $T$-$P$ profile. 
In Fig \ref{fig:abundance_T}, we show the computed abundances at 1 bar level as a function of $T_{\rm eq}$. From the figure, CO, CO$_2$, and N$_2$ abundances increase as $T_{\rm eq}$ increases, while CH$_4$, NH$_3$, and PH$_3$ abundances decrease as $T_{\rm eq}$ increases. H$_2$O and H$_2$S abundances remain approximately unchanged relative to the change of $T_{\rm eq}$. The HCN abundance increases and decreases as $T_{\rm eq}$ increases. 

\section{Results for noiseless spectra modeling}\label{sec:noiseless_spectra}

In this section, we present the synthetic primary and secondary transit spectra for three fiducial planets ($T_{\rm eq}$ = 500 K, 1000 K, and 1500 K). The parameters for the planets are summarized in Table \ref{tab:system_params}. The planets are solar composition Jupiter-size planets with different levels of insolation. The vertical $T$-$P$ profiles for the planets are presented in Fig. \ref{fig:PT_profile} and the vertical abundance profiles are presented in Fig \ref{fig:major_species}. The molecules included are H$_2$O, CO, CH$_4$, CO$_2$, N$_2$, NH$_3$, HCN, H$_2$S and PH$_3$.  

The spectral features for H$_2$O, CO, CH$_4$, CO$_2$, and NH$_3$ have been explored in the literature \citep[e.g.][]{Greene16}. From our calculations in section \ref{sec:abundances}, we found that H$_2$S is the primary carrier of sulfur for all different equilibrium temperatures; PH$_3$ is the primary carrier of phosphorus for $T_{\rm eq}$ $<$ 1000 K; HCN has a mixing ratio of 1 ppm for solar composition atmospheres with $T_{\rm eq}$ = 1000 K. These molecules are potentially identifiable from the transit spectra. Although HCN is not the primary carrier of either carbon or nitrogen, it is a disequilibrium species and its abundances are indicative of the strength of vertical mixing. Therefore, we also investigate the spectral feature of HCN and see if JWST can potentially detect HCN. 

Here we focus on identifying spectral features for PH$_3$, H$_2$S, and HCN in the primary and secondary transit spectra of solar composition Jupiter-size planets. We compare the spectra including all nine species with the spectra with one specific species excluded, in order to find the spectral feature of that specific species. 

\subsection{PH$_3$}\label{subsec:clean_PH$_3$}
\begin{figure*}
\gridline{\fig{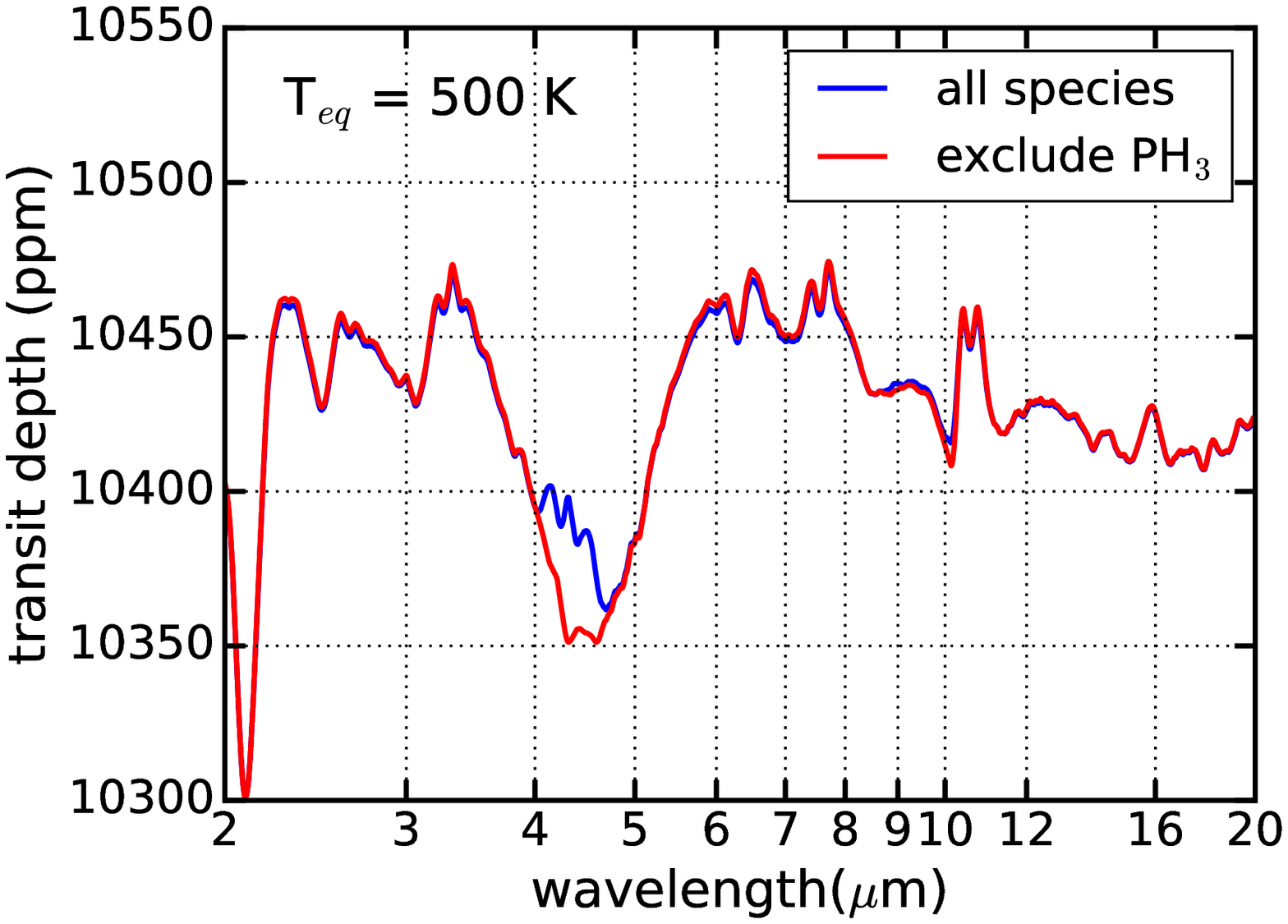}{0.5\textwidth}{(a)}
          \fig{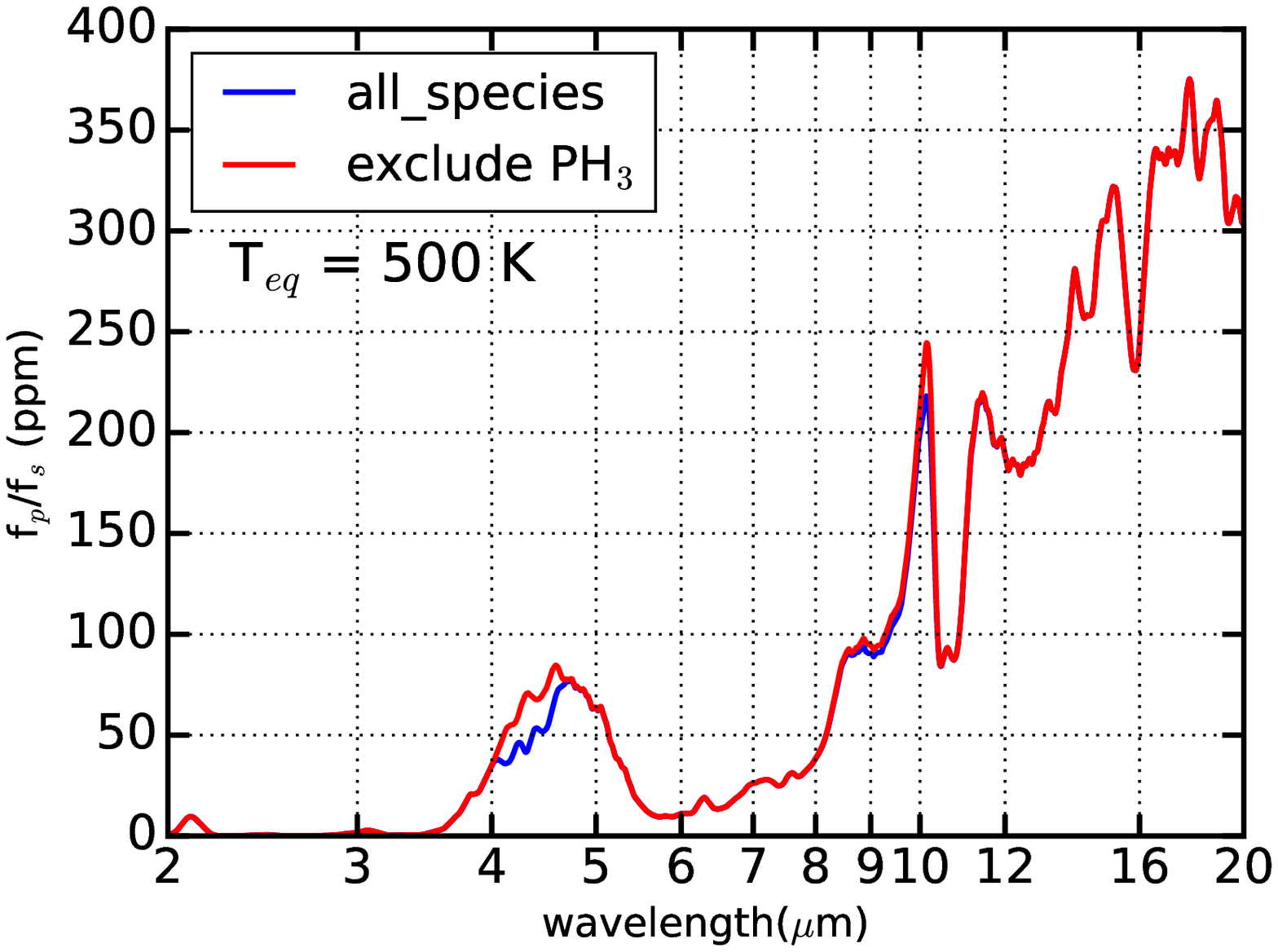}{0.5\textwidth}{(b)}
          }
\gridline{\fig{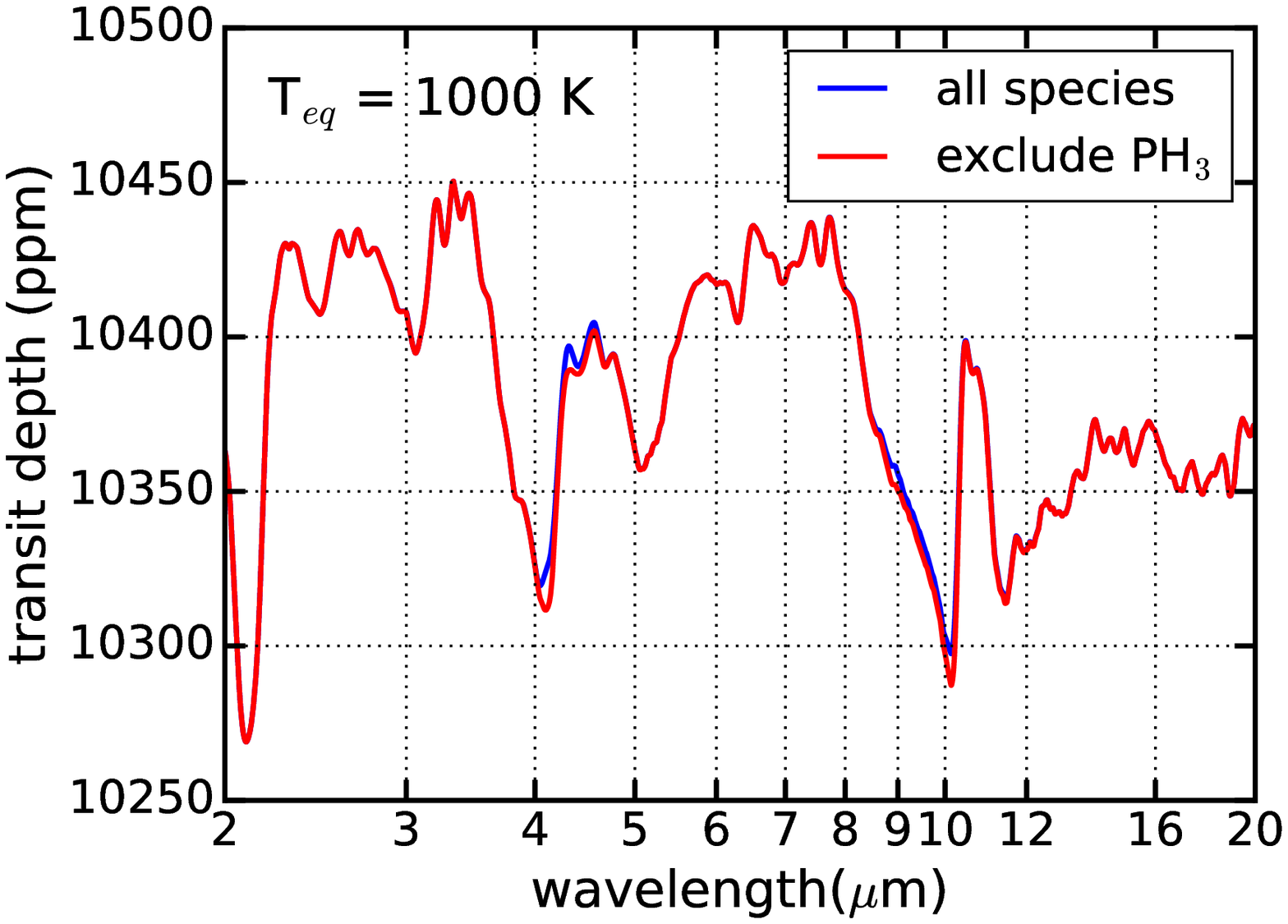}{0.5\textwidth}{(c)}
          \fig{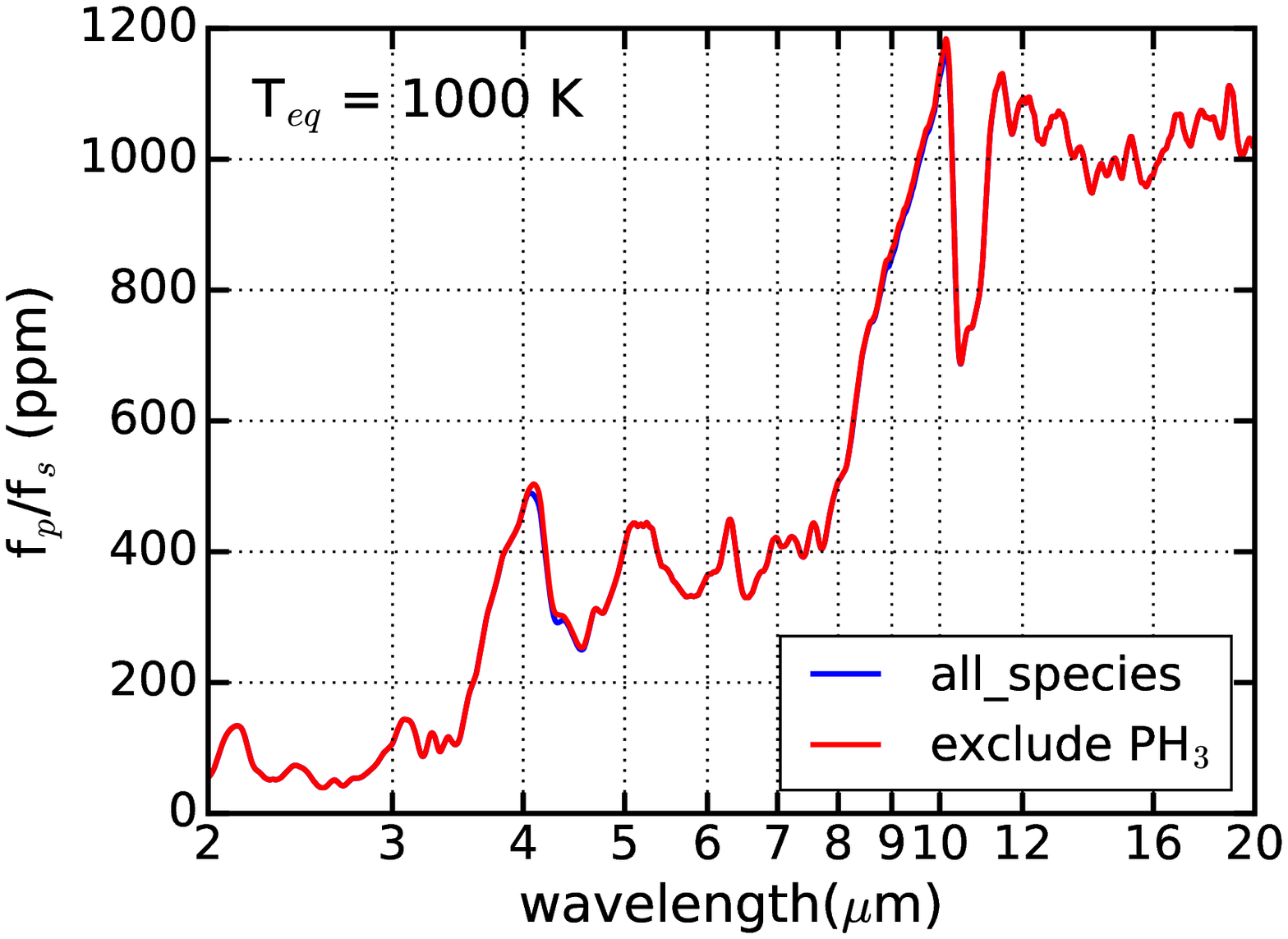}{0.5\textwidth}{(d)}
          }
\gridline{\fig{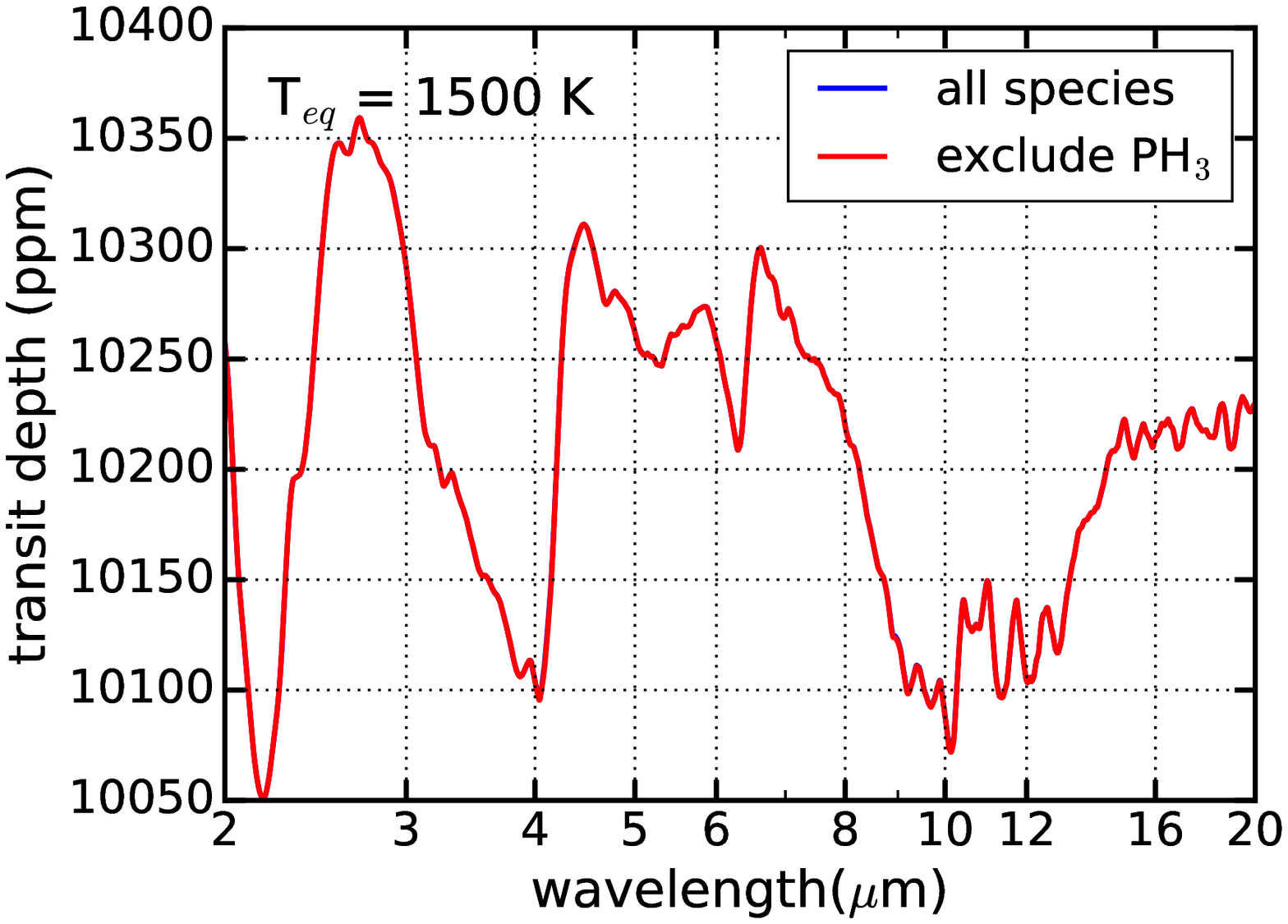}{0.5\textwidth}{(e)}
          \fig{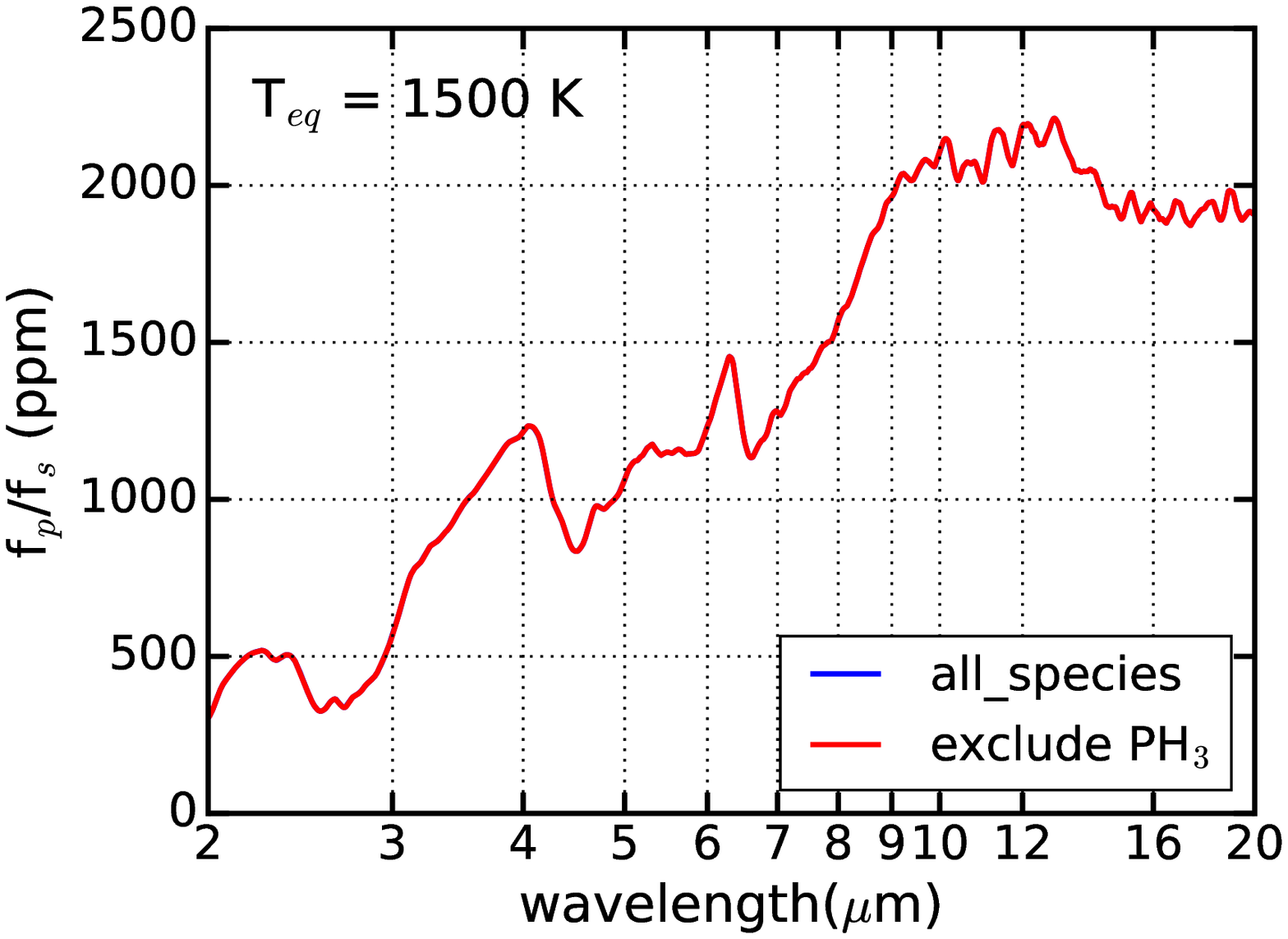}{0.5\textwidth}{(f)}
          }
\caption{Simulated transmission and emission spectra for \textit{all species} (including H$_2$O, CO, CH$_4$, CO$_2$, NH$_3$, N$_2$, HCN, H$_2$S, and PH$_3$) compared with \textit{all species except} PH$_3$. The difference between the blue curve and the red curve indicates the absorption from PH$_3$. The spectra are smoothed to a resolution of 100. 
\label{fig:clean_PH3_spectra}}
\end{figure*}

Fig. \ref{fig:clean_PH3_spectra}, we present the primary and secondary transit spectra for the planetary systems in Table \ref{tab:system_params}, with $T_{\rm eq}$ = 500 K, 1000 K, and 1500 K. The planets are assumed to have solar composition atmospheres. The vertical abundance profiles are taken from Fig. \ref{fig:major_species}. We compare the spectra simulated with \textit{all species}, and the spectra simulated with \textit{all species except} PH$_3$. The difference between these two spectra indicates the absorption from PH$_3$. For an solar composition atmosphere with $T_{\rm eq}$ = 500 K, the absorption from PH$_3$ occurs between 4 and 5 $\micron$. The absorption depth is about 40 ppm in the primary transit spectra. The absorption is about 20 ppm in the secondary transit spectra. For the atmosphere with $T_{\rm eq} = 1000 K$, the absorption is about 5 ppm in the primary transit spectra, while in the secondary transit spectra, the absorption is too small to be seen in the figure. For the atmosphere with $T_{\rm eq} = 1500 K$, there is no apparent PH$_3$ absorption feature in the spectra.

The lack of PH$_3$ spectral feature for the solar composition atmospheres with $T_{\rm eq}$ = 1000 K and 1500 K is due to the thermal decomposition of PH$_3$ under high temperatures. From Fig. \ref{fig:phosphorus_chemistry}, for the atmospheres with $T_{\rm eq} = 500 K$, almost all of the phosphorus are in the form of PH$_3$; while for the atmospheres with $T_{\rm eq}$ = 1000 K and 1500 K, most phosphorus are in the form of P$_2$ and PH$_2$. Therefore, the spectral features of PH$_3$ are only expected in moderately irradiated atmospheres.    

\subsection{H$_2$S}\label{subsec:clean_H_2S}
\begin{figure*}
\gridline{\fig{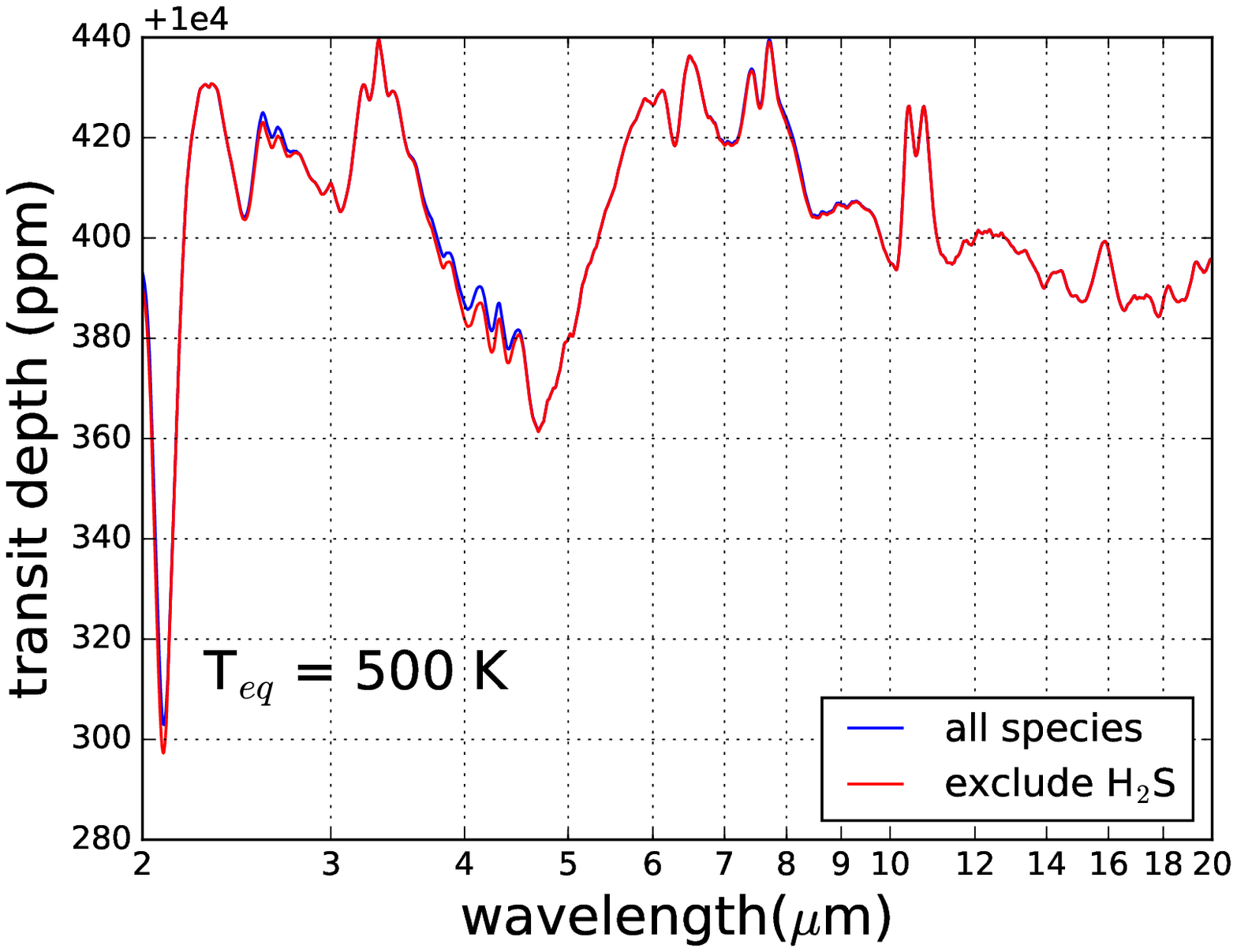}{0.5\textwidth}{(a)}
          \fig{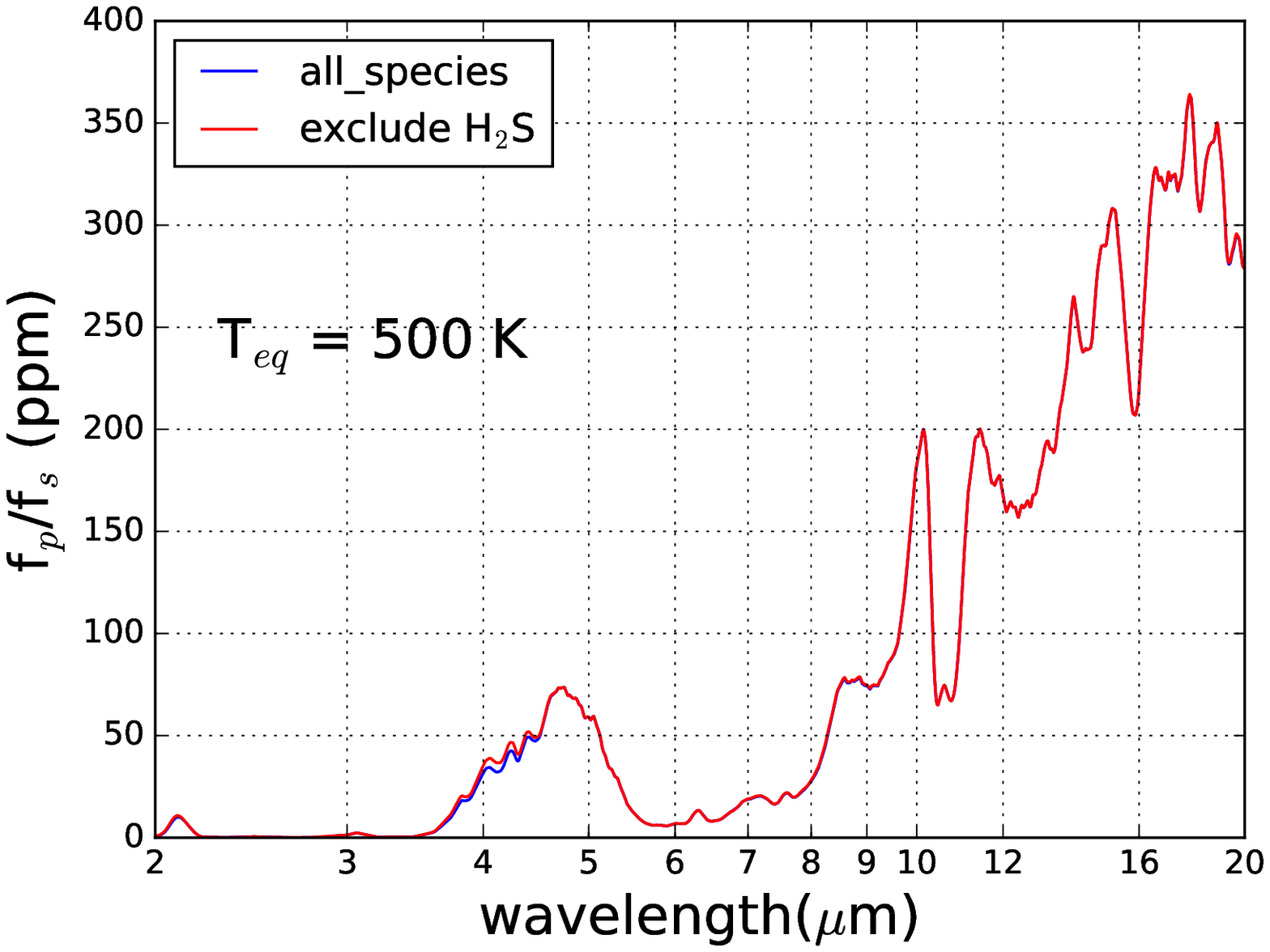}{0.5\textwidth}{(b)}
          }
\gridline{\fig{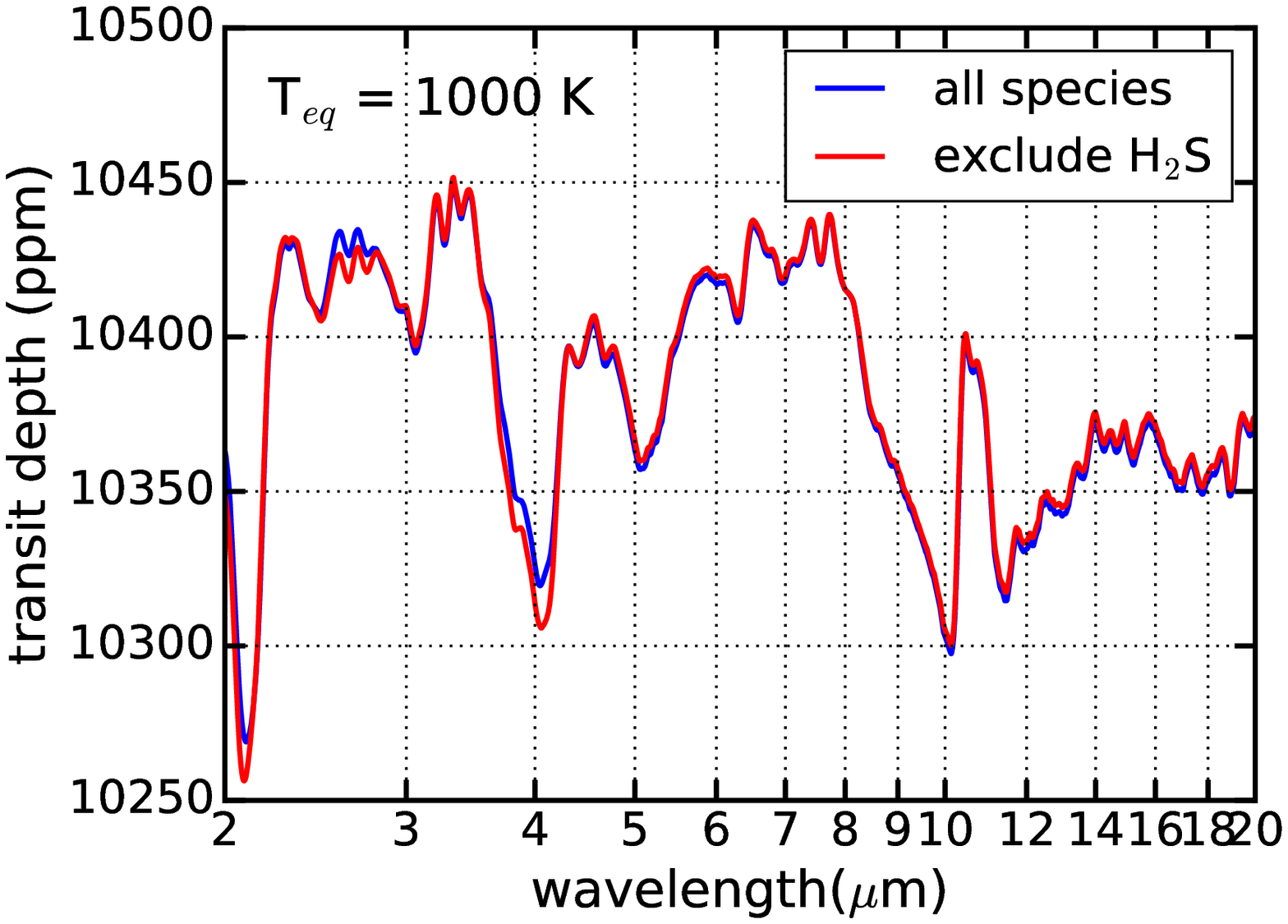}{0.5\textwidth}{(c)}
          \fig{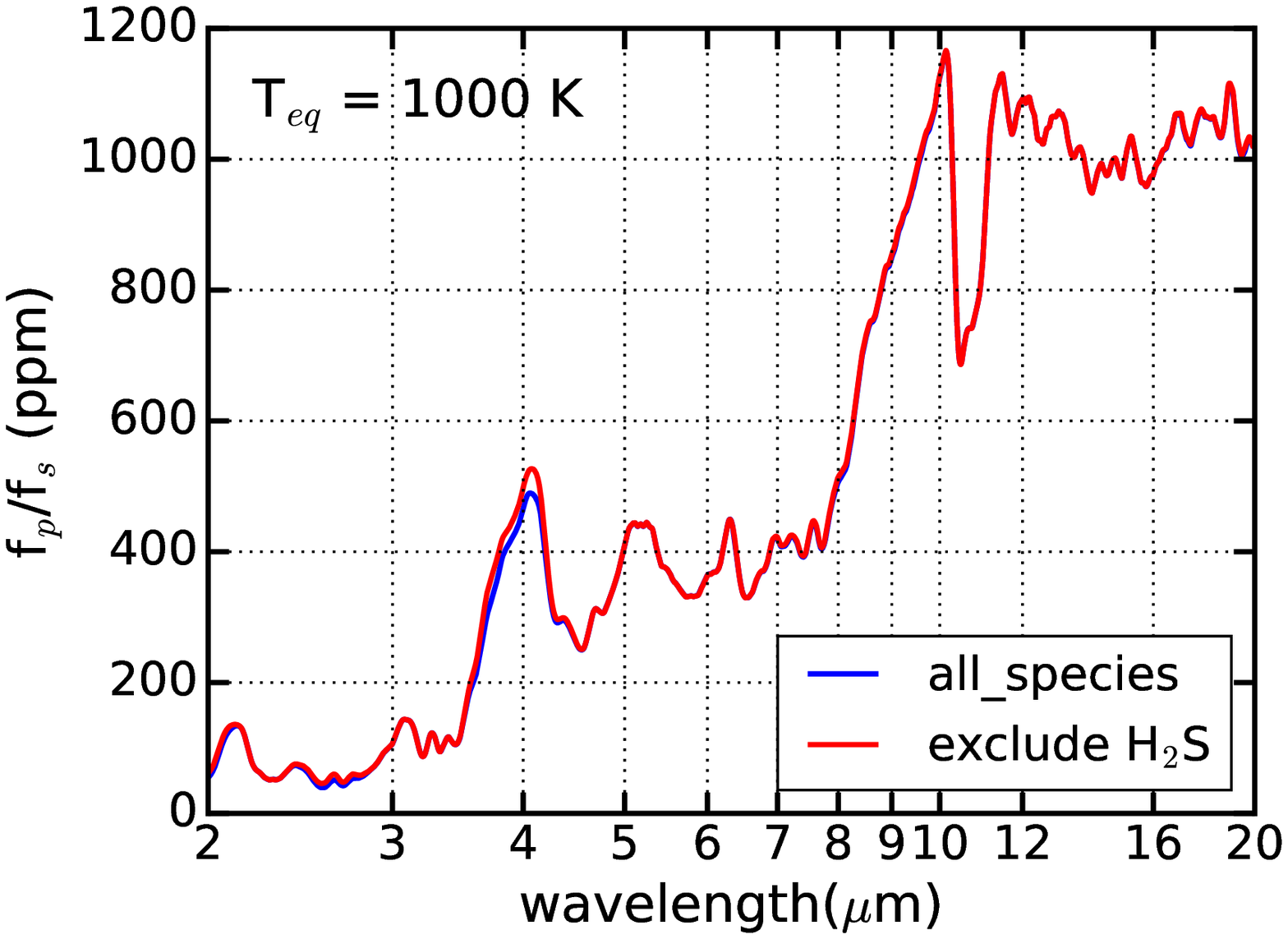}{0.5\textwidth}{(d)}
          }
\gridline{\fig{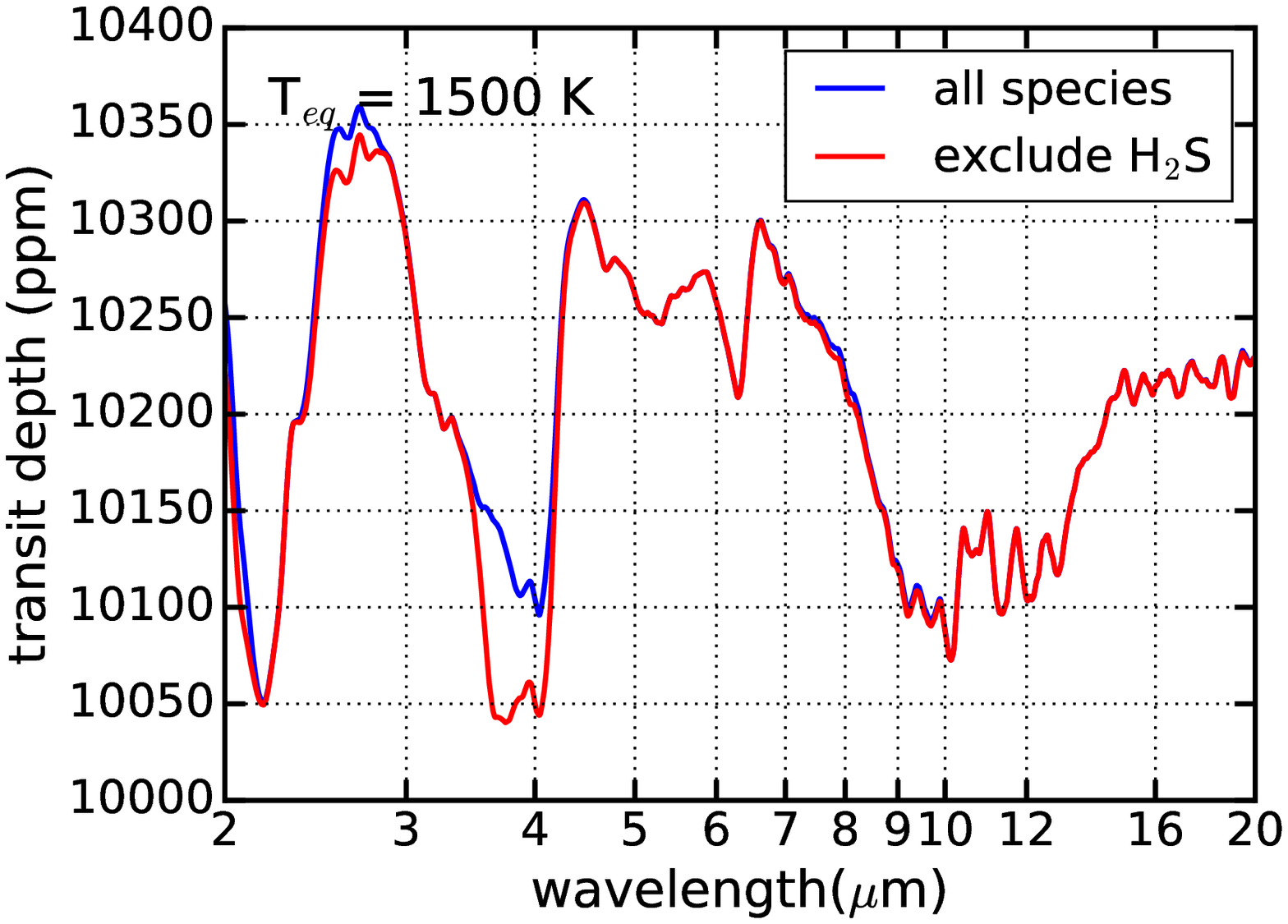}{0.5\textwidth}{(e)}
          \fig{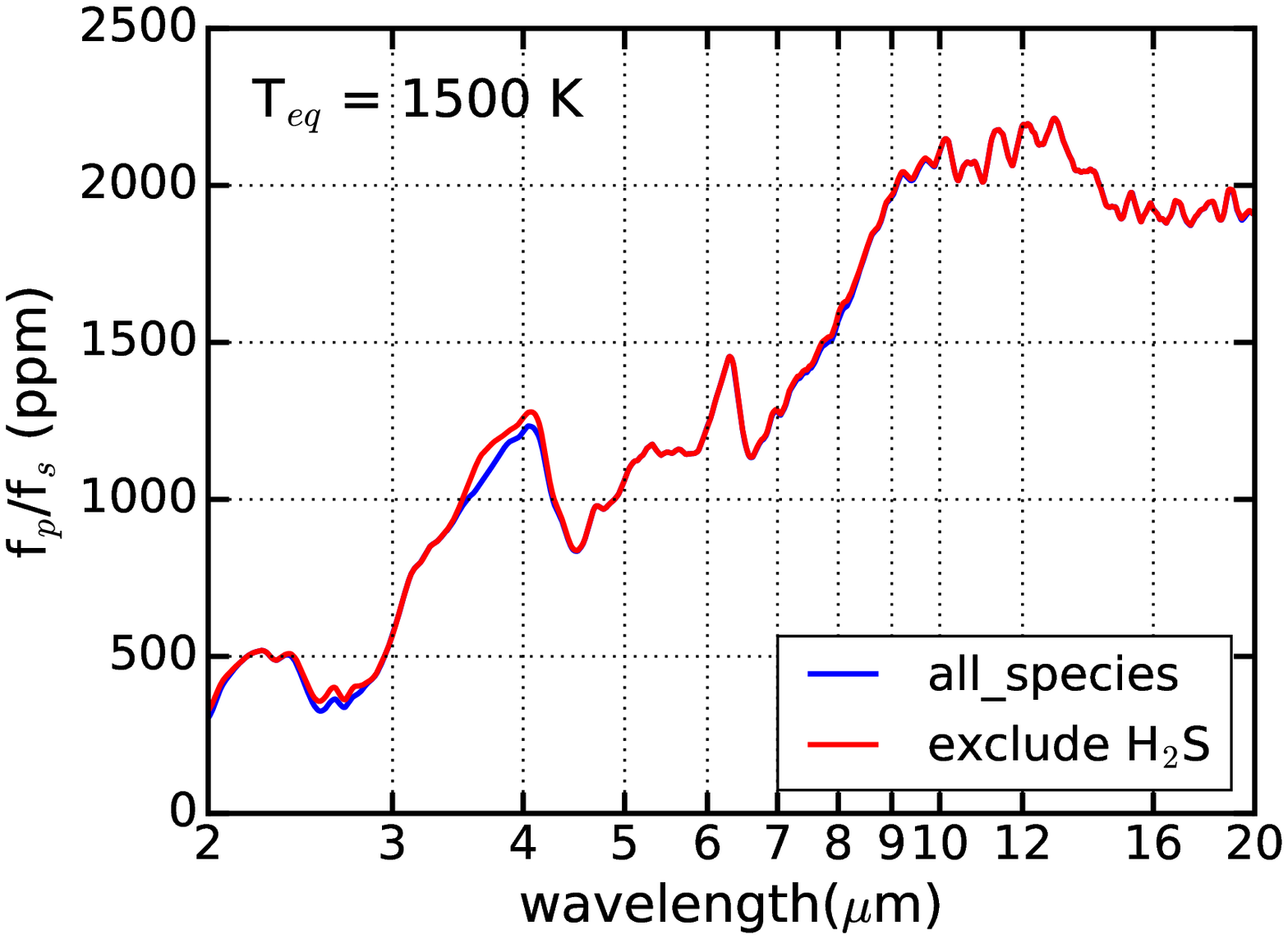}{0.5\textwidth}{(f)}
          }
\caption{Simulated transmission and emission spectra for \textit{all species} (including H$_2$O, CO, CH$_4$, CO$_2$, NH$_3$, N$_2$, HCN, H$_2$S, and PH$_3$) and \textit{all species except} H$_2$S. The difference between the red curve and the blue curve indicates the absorption by H$_2$S. The spectra is smoothed to a resolution of 100.
\label{fig:clean_H2S_spectra}}
\end{figure*}

In Fig. \ref{fig:clean_H2S_spectra}, we present the synthetic primary and secondary transit spectra for the solar composition atmospheres with $T_{\rm eq}$ = 500 K, 1000 K, and 1500 K in Table \ref{tab:system_params}. We compare the spectra simulated with \textit{all species}  and the spectra simulated with \textit{all species except} H$_2$S. For the atmospheres with $T_{\rm eq}$ = 500 K, the absorption depth is very small. In the primary transit spectra, there is a 5 ppm absorption at 2.6 $\sim$ 2.8 $\micron$ and a 10 ppm absorption at 3.9 $\sim$ 4.3 $\micron$. In the secondary transit spectra, there is a 10 ppm absorption at 3.9$\sim$ 4.3 $\micron$. For the atmospheres with $T_{\rm eq}$ = 1000 K, the absorption depths are also very small, for both primary transit and secondary transit spectra. For the atmospheres with $T_{\rm eq}$ = 1500 K, the absorption depths are much bigger. In the primary transit spectra, the absorption depth is about 15 ppm at 2.6$\sim$2.8 $\micron$, and 100 ppm at 3.5$\sim$4.1 $\micron$. In the secondary spectra, the absorption depth is about 10 ppm at 2.6 $\sim$ 2.8 ppm and 100 ppm at 3.5$\sim$ 4.1 $\micron$.   

The spectral feature of H$_2$S is more prominent in highly irradiated atmospheres. What determines the relevance of H$_2$S in the spectra is other species. In cold atmospheres, H$_2$S has to compete with the more abundant NH$_3$ and CH$_4$ to absorb photons while in the hottest case those two molecules are less abundant, leaving more space to H$_2$S to absorb photons and be seen in the spectra. Another factor that may also contribute is the larger pressure scale height in hotter atmospheres. 

\subsection{HCN}\label{subsec:clean_HCN}

\begin{figure*}
\gridline{\fig{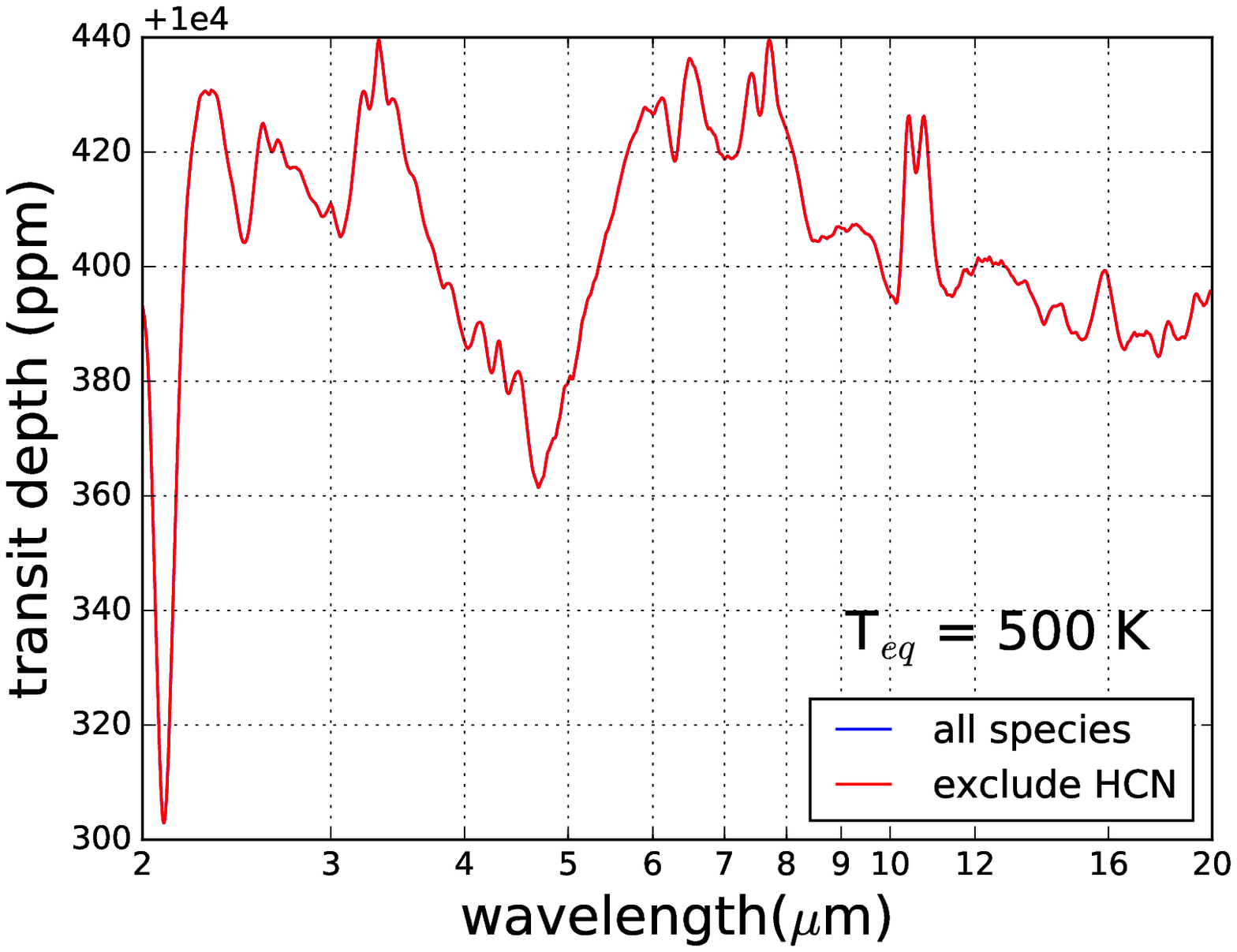}{0.5\textwidth}{(a)}
          \fig{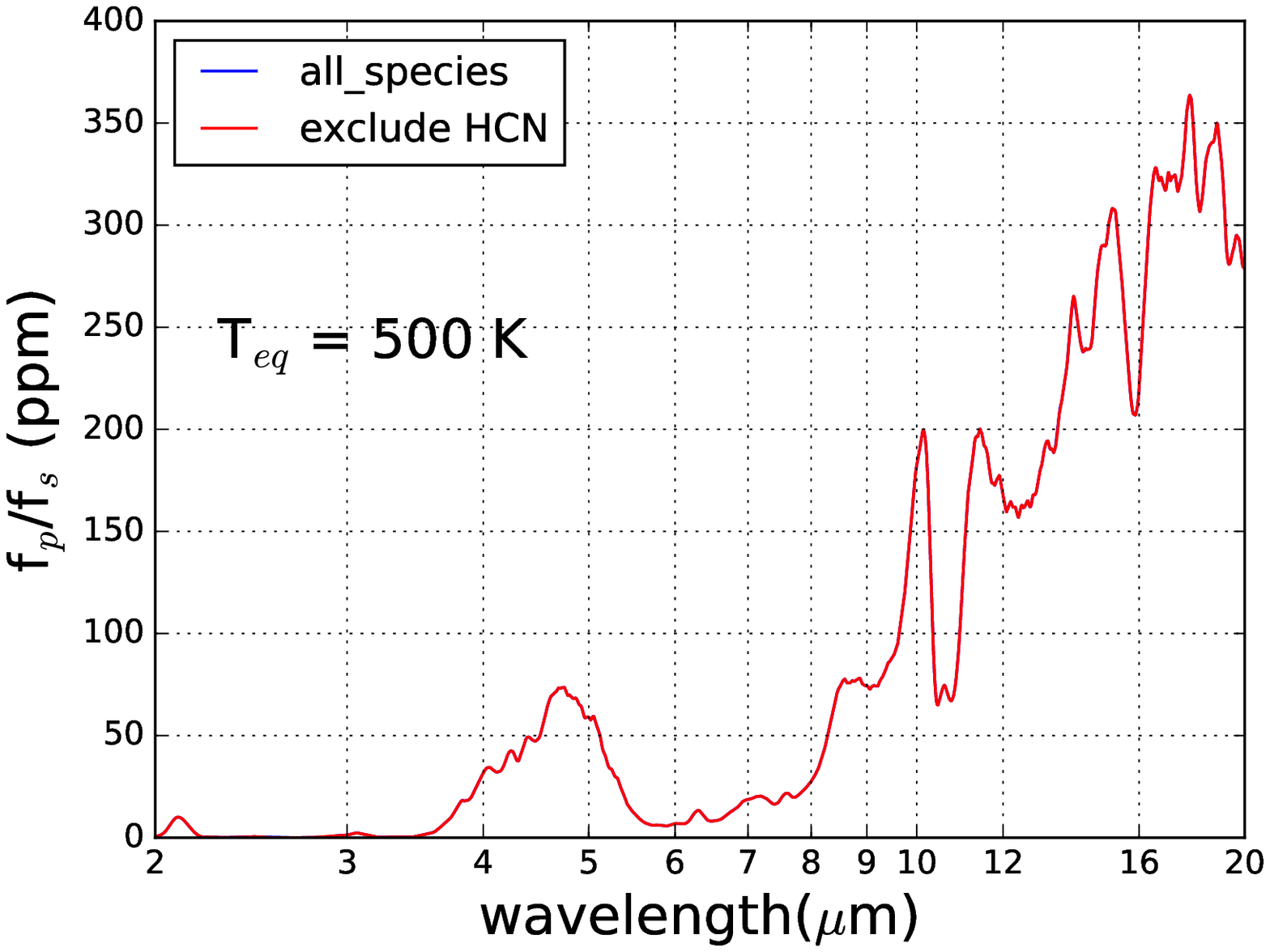}{0.5\textwidth}{(b)}
          }
\gridline{\fig{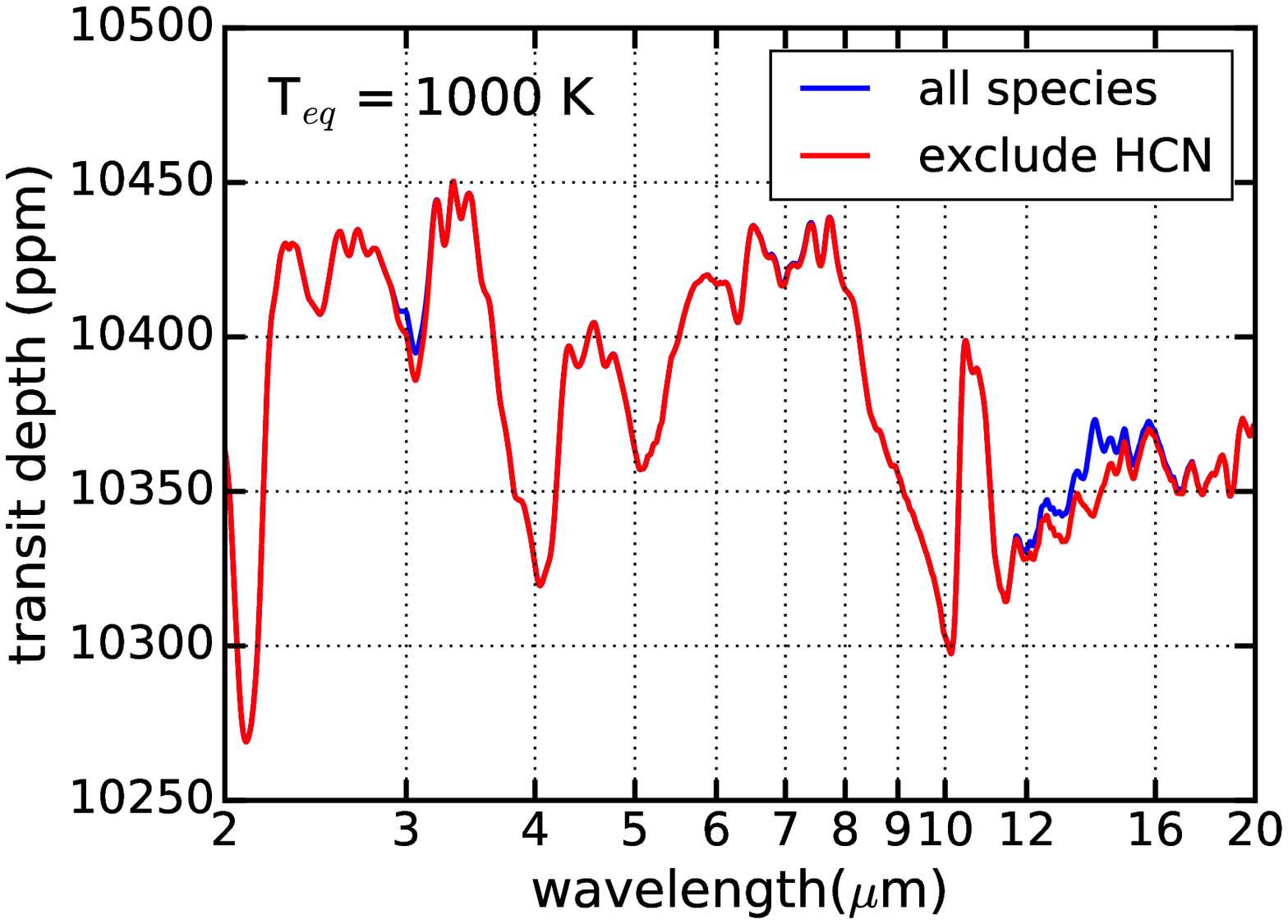}{0.5\textwidth}{(c)}
          \fig{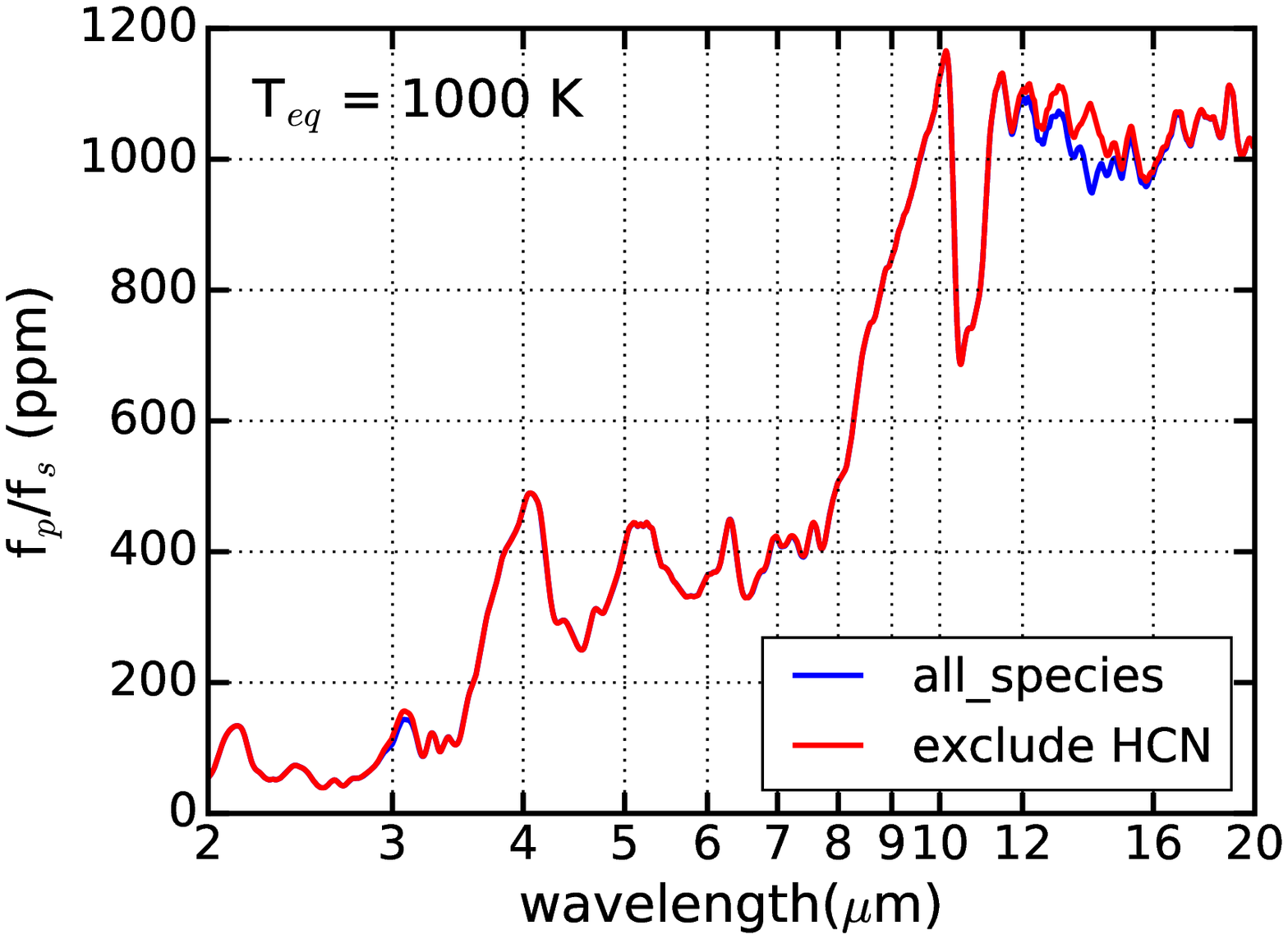}{0.5\textwidth}{(d)}
          }
\gridline{\fig{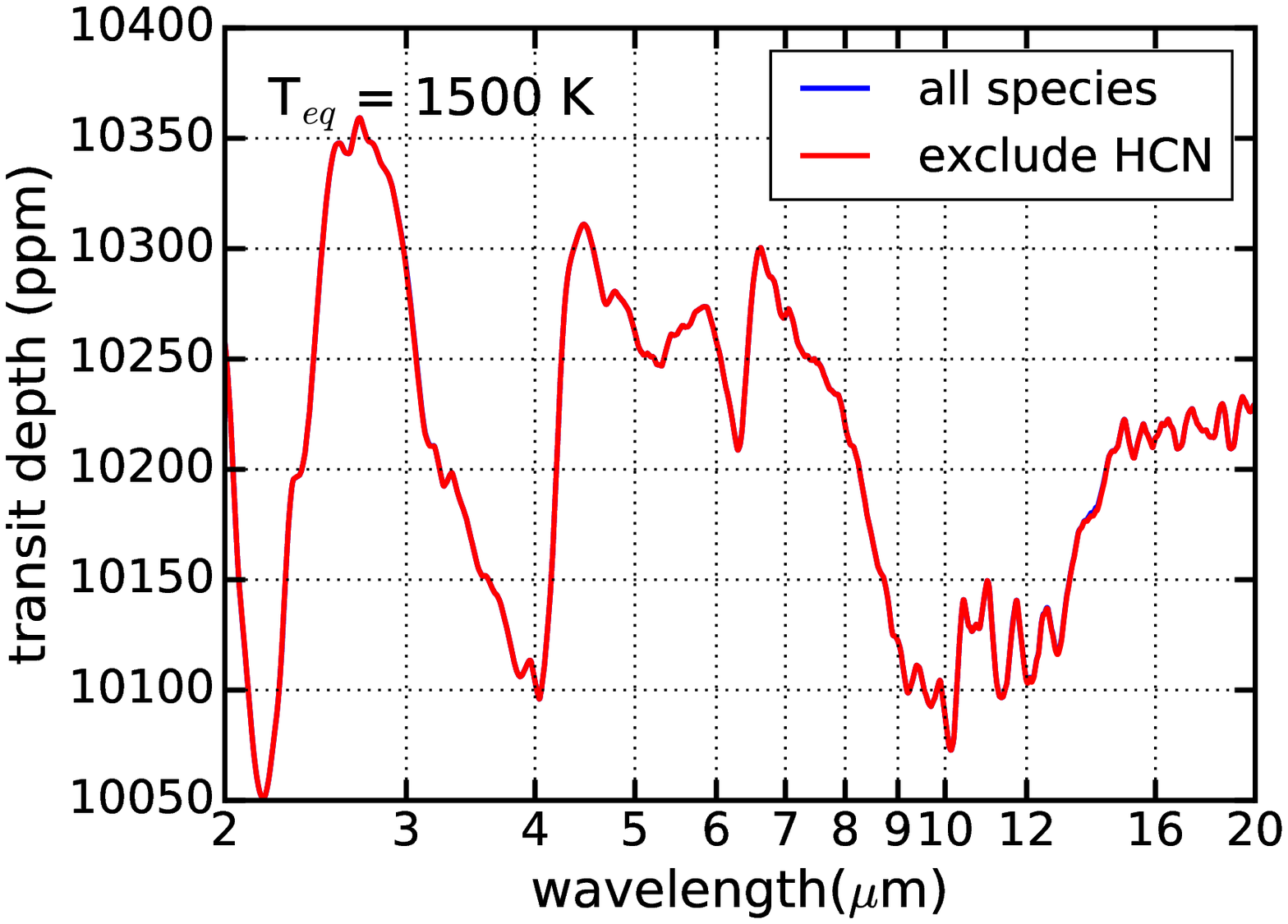}{0.5\textwidth}{(e)}
          \fig{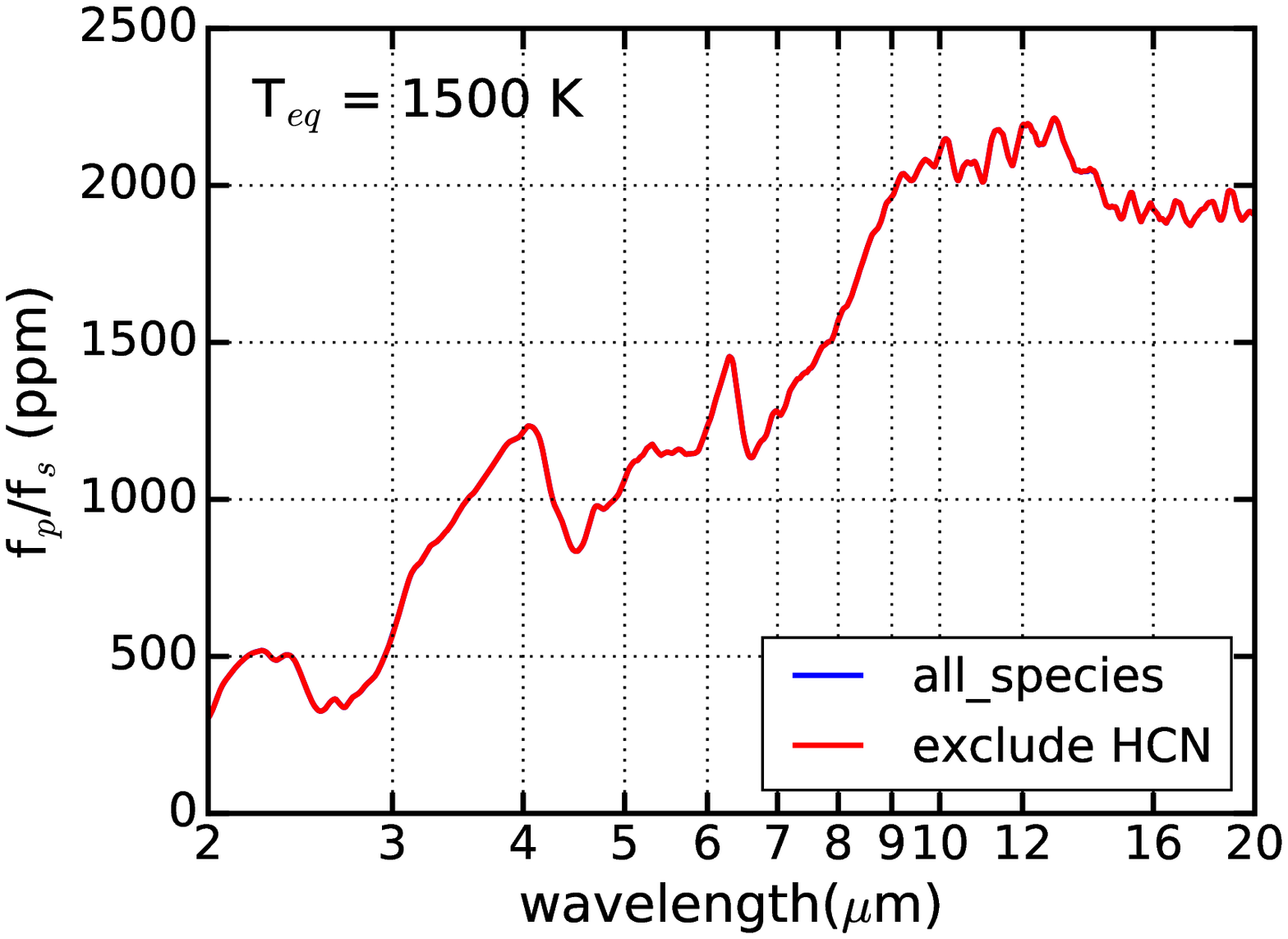}{0.5\textwidth}{(f)}
          }
\caption{Simulated transmission and emission spectra for \textit{all species} (H$_2$O, CO, CH$_4$, CO$_2$, NH$_3$, N$_2$, HCN, H$_2$S, and PH$_3$) and \textit{all species except} HCN. The difference between the red curve and the blue curve indicates the absorption by HCN. The spectra is smoothed to a resolution of 100.
\label{fig:clean_HCN_spectra}}
\end{figure*}

In Fig. \ref{fig:clean_HCN_spectra}, we present the synthetic primary and secondary transit spectra for solar composition atmospheres with $T_{\rm eq}$ = 500 K, 1000 K, and 1500 K listed in Table \ref{tab:system_params}. For the atmospheres with $T_{\rm eq}$ = 500 K and 1500 K, there are little absorption from HCN, mainly because the mixing ratio of HCN is very low, as shown in Fig. \ref{fig:major_species}. For the atmospheres with $T_{\rm eq}$ = 1000 K, there are small absorption features between 12 and 16 $\micron$. The absorption depth in the primary transit spectra is about 15 ppm, and the absorption depth in the secondary transit spectra is about 80 ppm. 

\section{Results for JWST transit spectra modeling}\label{sec:noise_spectra}

In this section, we model the JWST spectra for primary and secondary transit observations . The instruments and modes for transit observations are shown in Table \ref{tab:noise_params}. The wavelength range modeled is between 2.5 $\micron$ and 11 $\micron$. In section \ref{sec:noiseless_spectra}, we identified spectral features for PH$_3$, H$_2$S, and HCN. The spectral feature of PH$_3$ is between 4 $\micron$ and 5 $\micron$, the spectral feature of H$_2$S is between 3 $\micron$ and 4 $\micron$, and the spectral feature of HCN is between 12 $\micron$ and 16 $\micron$. The feature of HCN is beyond the limit of MIRI LRS mode \citep{Beichman14}. In this paper, we focus on the spectra features of H$_2$S and PH$_3$. 

\subsection{Results for JWST noise modeling}

\begin{figure*}
\plotone{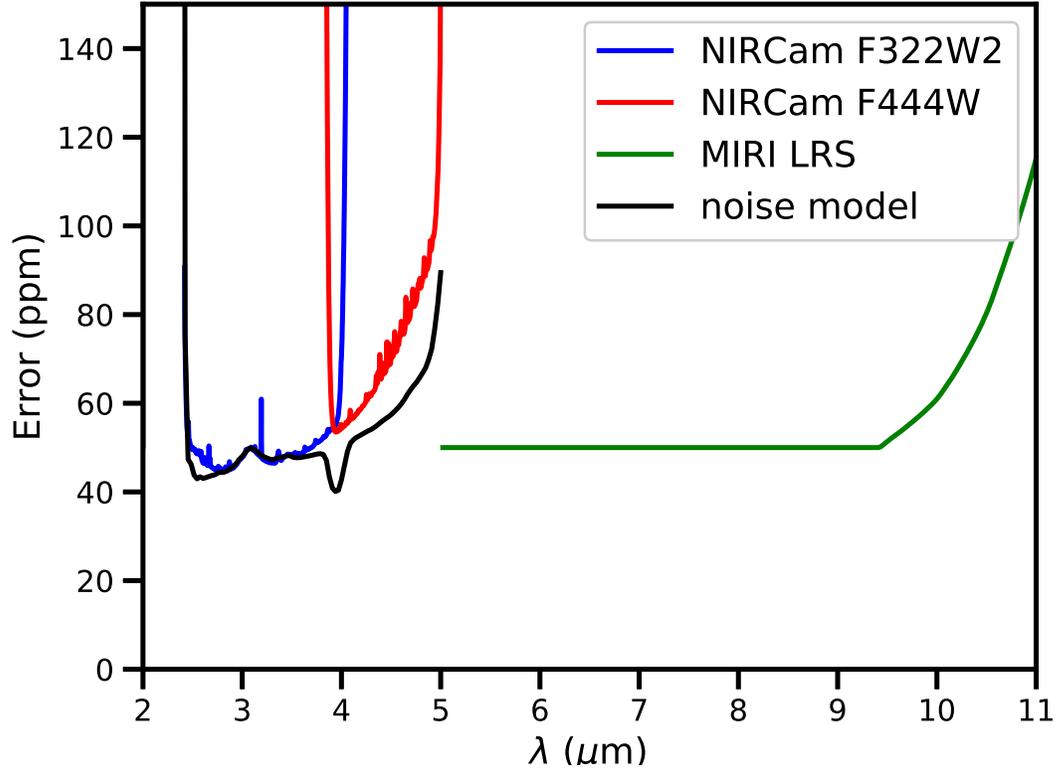}
\caption{Noise level for different observing modes as calculated using PandExo. The target is K = 6.8 G type star. The transit duration for this calculation is 7.2 hrs. As a comparison, we also plot the error level as calculated following the recipe in \citet{Greene16}.    
\label{fig:noise_lambda}}
\end{figure*}

We used both PandExo and the Greene et al. (2016) noise model to compute the error on the spectrum. In Fig. \ref{fig:noise_lambda}, we show the noise level as a function of wavelength for a range of instrument modes covering wavelength ranging from 2.5 $\micron$ to 11 $\micron$. The parameters of the planetary system being modeled are listed in Table \ref{tab:system_params}. We assume two transits, each with equal in-transit and out-transit integration time. The transit duration is 7.2 hrs, and the total (transit + baseline) time is 14.4 hrs. The number of groups and number of integrations are optimized by the PandExo. Besides the results calculated using PandExo, we also compute the error using the Greene et al. (2016) noise model for the NIRCam F322W2 mode and F444W mode. The Greene et al. (2016) model gives slightly lower noise levels, but similar to the results by PandExo.

\subsection{PH$_3$}
\begin{figure*}
\gridline{\fig{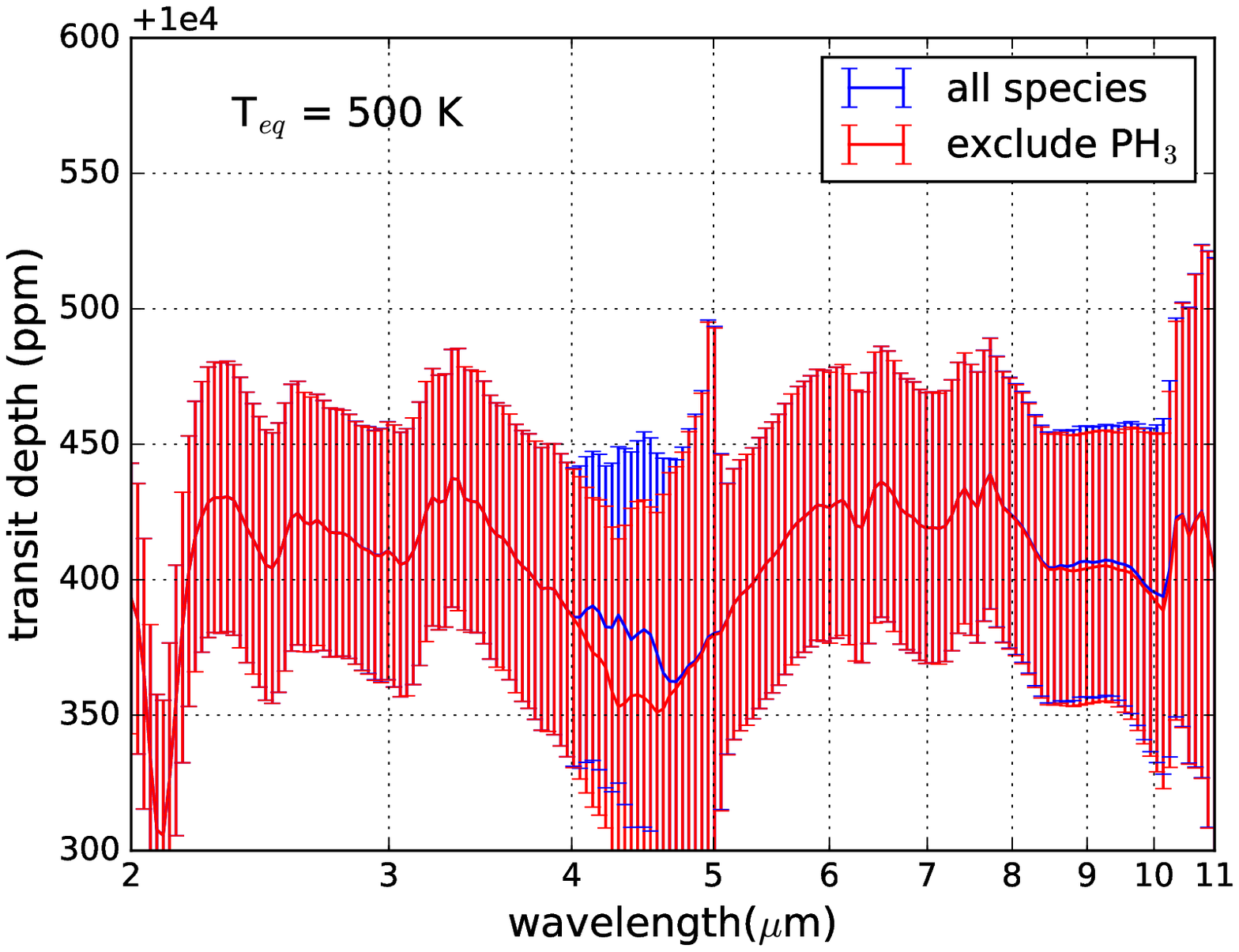}{0.5\textwidth}{(a)}
          \fig{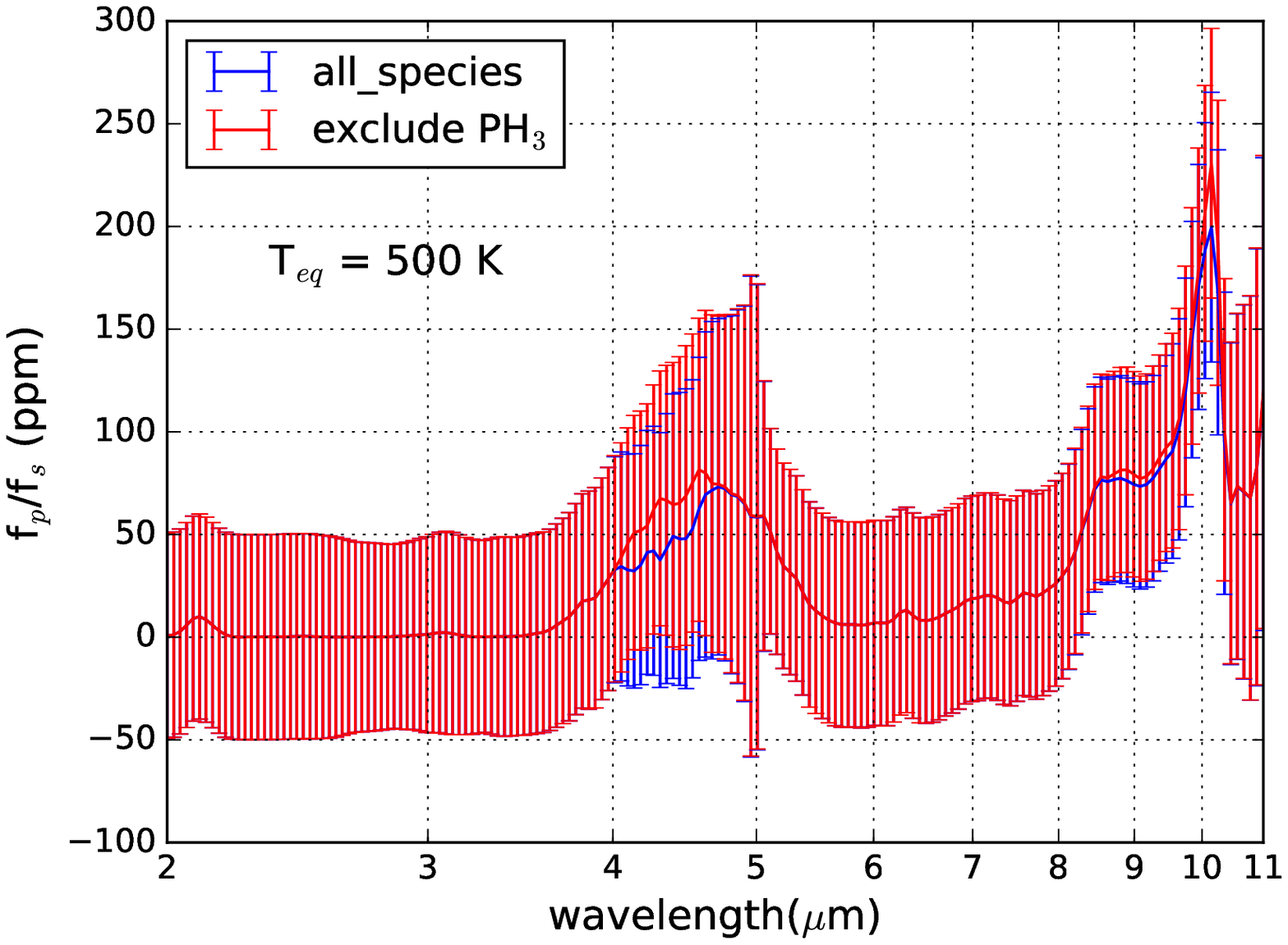}{0.5\textwidth}{(b)}
          }
\caption{Synthetic transmission and emission spectra with simulated JWST noise for the planet presented in Table \ref{tab:system_params} with $T_{\rm eq}$ = 500 K. The blue curve is the binned spectra simulated including all nine species, and the red curve is the binned spectra simulated including all nine species except PH$_3$. The error bars in the plot indicate one sigma Gaussian noise level simulated for JWST instrument modes summarized in Table \ref{tab:noise_params}. The total required observing time is 28.8 hrs.
\label{fig:PH3_spectra}}
\end{figure*}

In Fig. \ref{fig:PH3_spectra}, we show the synthetic primary and secondary transit spectra with simulated JWST noise and compare the spectra with and without PH$_3$. The planetary system being modeled is listed in Table \ref{tab:system_params} with $T_{\rm eq}$ = 500 K. The observation parameters are summarized in Table \ref{tab:observing_500K}. The spectral absorption feature of PH$_3$ is between 4.0 and 4.7 $\micron$. In the primary transit spectra, the absorption depth for PH$_3$ is approximately one standard deviation of the noise. Since there are $\sim$ 10 measurements within this feature, the significance level for detecting PH$_3$ is more than three sigma.  In order to cover the 4.0 -- 4.7 $\micron$ wavelength range, the NIRCam LW grism mode with F444W filter can be used in the observation. To achieve this level of error, two transits are needed. The integration time for each transit is about 7.2 hrs. We assume equal integration time for in transit and out of transit, and that leads to a total observing time of 7.2 hrs $\times$ 2 in and out of transit integration $\times$ 2 transits = 28.8 hrs.  
The absorption feature in the secondary transit spectra is harder to detect since the absorption depth from PH$_3$ is only about half the standard deviation of the noise. Therefore, it is the most effective for detecting PH$_3$ to use the primary transit spectra with the NIRCam LW grism mode and F444W filter, and get the spectra between 3.9 $\micron$ and 5.0 $\micron$. However, the feature is very weak and requires long integration time (7.2 hrs) to achieve the error level shown in Fig. \ref{fig:PH3_spectra}. Since CH$_4$ absorption is strong between 4 and 5 $\micron$, additional observation using the NIRCam LR F444W may be needed to break the degeneracy between the PH$_3$ absorption and CH$_4$ absorption. 
For higher equilibrium temperatures ($T_{\rm eq}$ $>$ 1000 K), the spectral feature of PH$_3$ is below 5 ppm that JWST is unlikely to detect it. 

\subsection{H$_2$S}
\begin{figure*}
\gridline{\fig{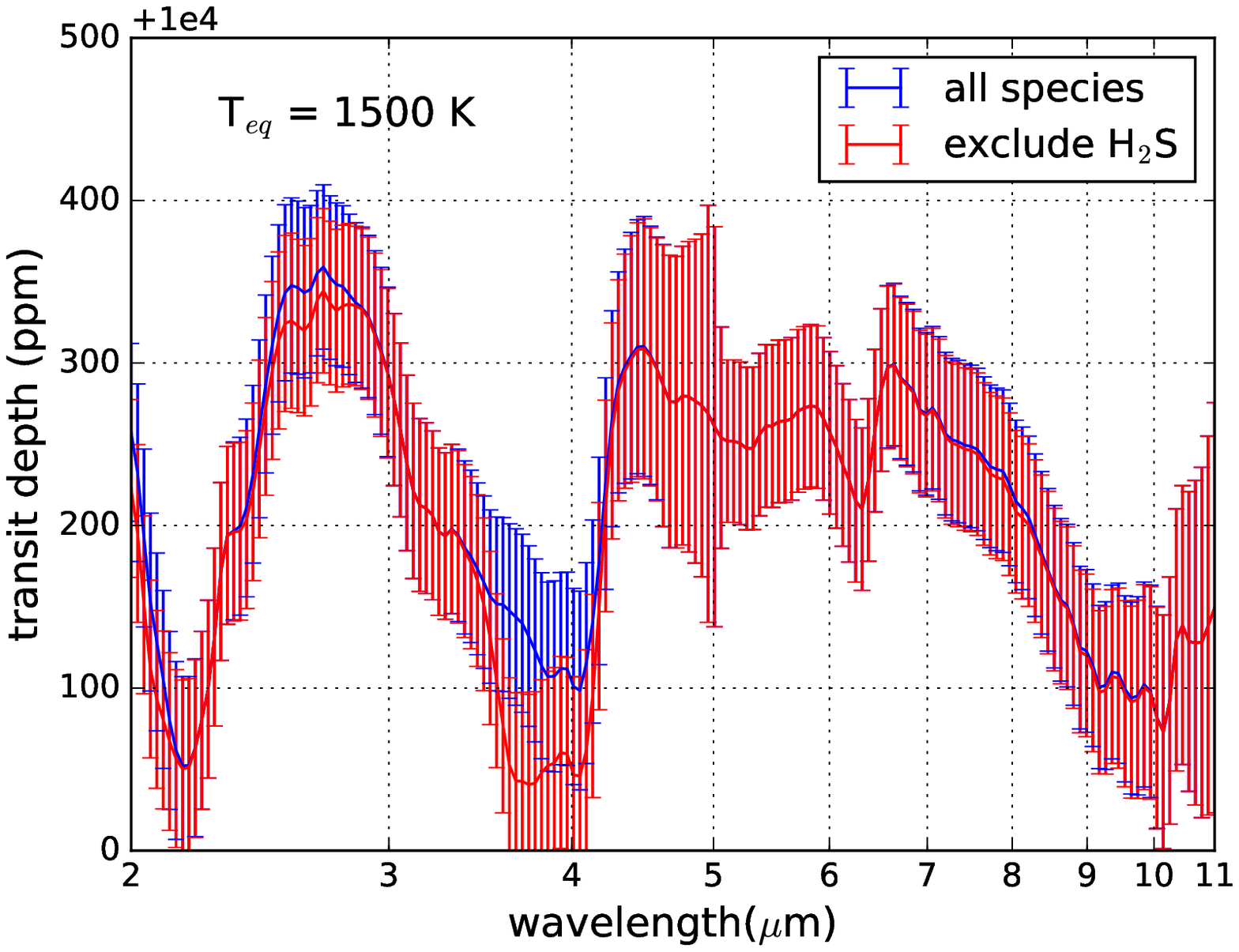}{0.5\textwidth}{(a)}
          \fig{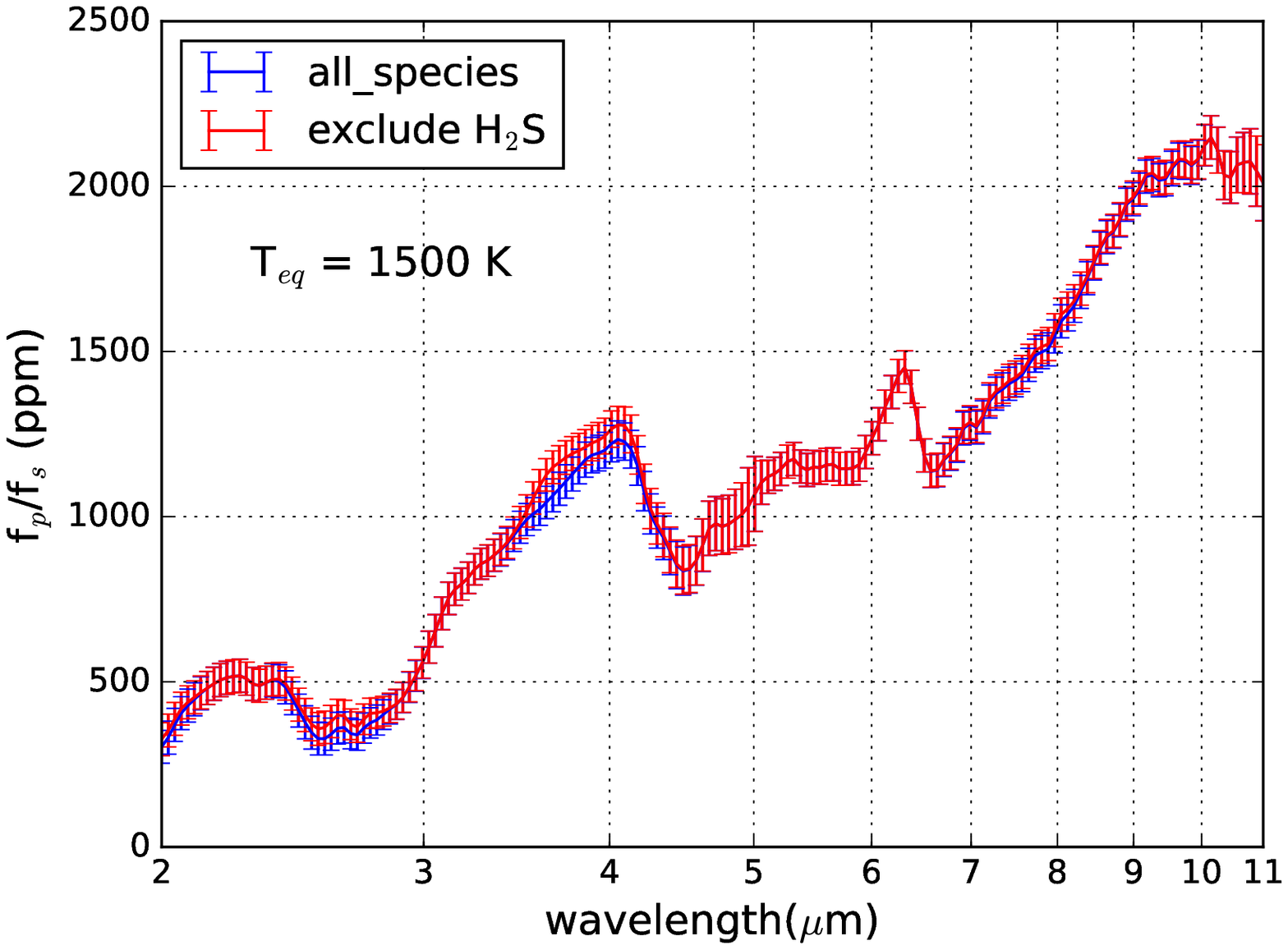}{0.5\textwidth}{(b)}
          }
\caption{Synthetic transmission and emission spectra with simulated JWST noise for planet presented in Table \ref{tab:system_params} with $T_{\rm eq}$ = 1500 K. The blue curve is the binned spectra simulated including all nine species in Fig. \ref{fig:major_species}, and the red curve is the binned spectra simulated including all nine species except H$_2$S. The error bars in the plot indicate one sigma Gaussian noise level simulated for JWST instrument modes summarized in Table \ref{tab:noise_params}. The total required observing time is 24.0 hrs.
\label{fig:H2S_spectra}}
\end{figure*}

In Fig. \ref{fig:H2S_spectra}, we show the synthetic primary and secondary transit spectra with simulated JWST noise and compare the spectra with and without H$_2$S. The spectral absorption feature of H$_2$S is between 3.5 and 4 $\micron$. We show the spectra with simulated noise for the planetary system in Table \ref{tab:system_params} with $T_{\rm eq}$ = 1500 K. The star in the system is a Sun-like star with K band magnitude of 6.8. The observational parameters are summarized in Table \ref{tab:observing_1500K}. In the primary transit spectra, the absorption depth is about two times the standard deviation of the noise. The shape of the absorption feature can be resolved with the binned spectral resolution of R $\sim$ 100. To cover the spectral feature of H$_2$S, one can use the NIRCam LW grism mode with F322W2 filter, getting the spectra between 2.4 $\micron$ and 4.0 $\micron$. In the secondary transit spectra, the absorption depth is approximately two times the standard deviation of the noise. Therefore, it is also likely to detect H$_2$S in the secondary transit spectra. The same mode of NIRCam can be used to obtain the secondary transit spectra. To achieve this level of error, five transits are needed. The integration time for each transit is about 2.4 hrs. We calculate the total observing time in the same way as in section 5.2, to obtain 24.0 hours.  Since H$_2$O absorption is also important in the 2.4 $\micron$ to 4.0 $\micron$ region, NIRISS SOSS mode observations may be needed to determine the base and peak of water features, in order to verify the spectra features of H$_2S$. We only show the case for $T_{\rm eq}$ = 1500 K in Fig. \ref{fig:H2S_spectra}, while for atmospheres with lower equilibrium equilibrium temperature (500 K, 750 K, 1000 K), the absorption from H$_2$S is below 10 ppm and JWST is unlikely to detect it.   

\section{Discussion}\label{sec:discussions}

In this paper, we investigate the phosphorus and sulfur chemistry in solar composition atmospheres of Jupiter-size planets with the effect of vertical mixing, but no photochemistry. We find PH$_3$ is the primary carrier of phosphorus in solar composition atmospheres with $T_{\rm eq}$ $\sim$ 500 K. For solar composition atmospheres with 1000 K $\lesssim$ $T_{\rm eq}$ $\lesssim$ 1500 K, the primary carrier of phosphorus is P$_2$. For very highly irradiated solar composition atmospheres ($T_{\rm eq}$ $\sim$ 2000 K), phosphorus is mainly sequestered in PH$_2$ and PH. For sulfur chemistry, we find H$_2$S is the primary carrier of sulfur for solar composition atmospheres with $T_{\rm eq}$ $<$ 2000 K. We also investigate the chemistry of C/N/O/S bearing species for solar composition atmospheres. The most abundant carbon and nitrogen bearing species depend on the level of insolation. For the atmosphere with $T_{\rm eq}$ $<$ 1000 K, H$_2$O, CH$_4$ and NH$_3$ are the most abundant molecules; while for the atmosphere with $T_{\rm eq}$ $>$ 1000 K, H$_2$O, CO, and N$_2$ are the most abundant molecules. With the computed abundance profiles for H$_2$O, CO, CO$_2$, CH$_4$, NH$_3$, N$_2$, H$_2$S, PH$_3$, and HCN, we model the synthetic primary and secondary transit spectra and identify spectral features for PH$_3$, H$_2$S, and HCN. The detectibility of PH$_3$ and H$_2$S with JWST transit observations are evaluated by simulating the noise levels. We find PH$_3$ can be detected in the primary transit spectra of solar composition atmospheres with $T_{\rm eq}$ $\lesssim$ 500 K using JWST NIRCam LW F444W mode, and the total observing time needs to be $\sim$28.8 hrs. H$_2$S can be detected in both primary and secondary spectra of solar composition Jupiter-size planets with $T_{\rm eq}$ $\gtrsim$ 1500 K using JWST NIRCam LW F322W2 mode with a total observing time of 24.0 hrs.

Our results imply that H$_2$S is an important absorber in the 3 - 4 $\micron$ region for solar composition atmospheres with $T_{\rm eq}$ $>$ 1500 K. Failure to include H$_2$S in the retrieval analysis of JWST spectra may lead to the non-fitting of the spectra, or more detrimentally, lead to wrong abundances of other molecules. The effect can only be determined by a full retrieval analysis, which is beyond the scope of this paper. however,we highlight the importance of H$_2$S in the 3 - 4 $\micron$ region of transit spectra. 
PH$_3$ show little features on the transit spectra of hot Jupiter's atmospheres ($T_{\rm eq}$ $>$ 1000 K), but show absorption features between 4 and 5 $\micron$ in moderately irradiated atmospheres ($T_{\rm eq}$ $\sim$ 500 K). Retrieval of molecular abundances for moderately irradiated planets needs to consider the absorption from PH$_3$. 

We also considered HCN in our model since we find HCN is non-negligible for planets with $T_{\rm eq}$ = 1000 K, with a mixing ratio of 1 ppm. Since photochemistry also produces HCN, we expect more HCN in the upper atmospheres. The absorption features of HCN are mainly between 12 and 16 $\micron$. This wavelength range is beyond the coverage of the MIRI LRS siltless mode. Therefore, we did not discuss further the detectability of HCN with JWST. 

Our modeling of JWST transit spectra makes several simplifications, and we discuss the influences below. 

\begin{figure*}
\plotone{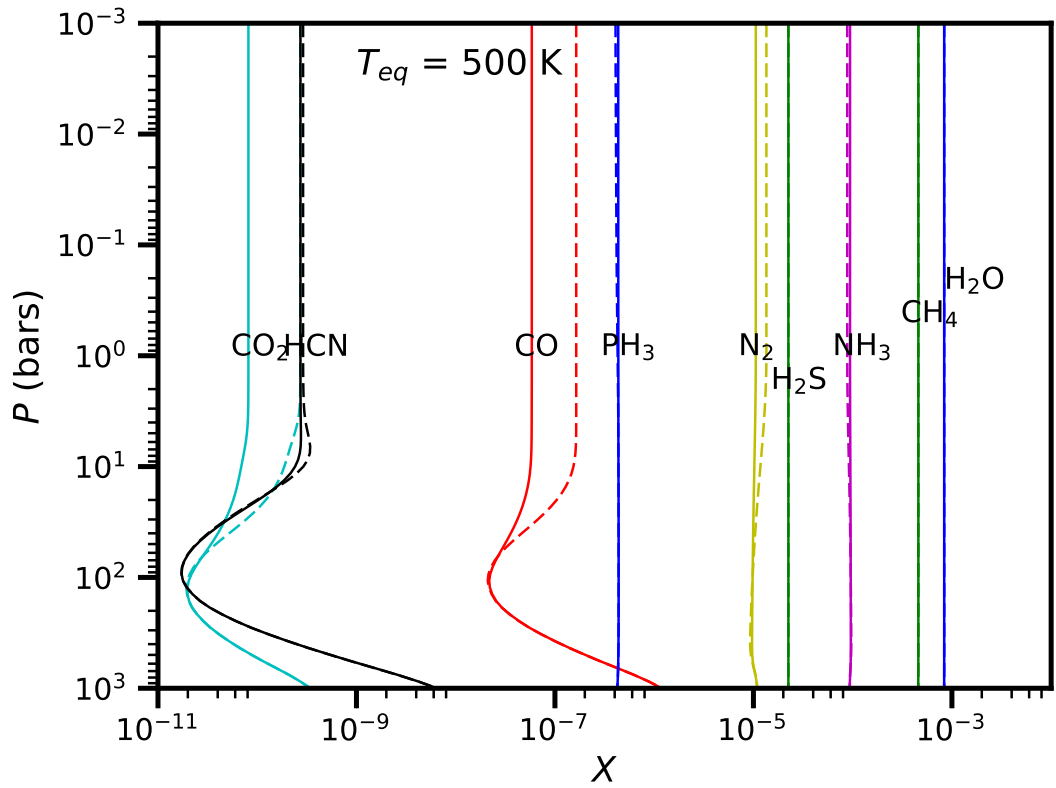}
\caption{Comparison on the vertical profiles of major molecules between two different levels of vertical mixing. Solid lines correspond to $K_{\rm eddy}$ = 1.0$\times$ 10$^9$ cm$^2$ s$^{-1}$, and dashed lines correspond to $K_{\rm eddy}$ =  1.0$\times$ 10$^8$ cm$^2$ s$^{-1}$. The planet being modeled has solar composition atmosphere with $T_{\rm eq}$ = 500 K. 
\label{fig:comparison_K}}
\end{figure*}

\begin{itemize}
\item The transit spectra of PH$_3$ and H$_2$S are complicated by the presence of clouds or hazes in the atmospheres of extrasolar giant planets. Current observations indicate that clouds or hazes are ubiquitous in the atmospheres of exoplanets \citep[e.g.][]{Pont08,Deming13,Kreidberg14a}. Clouds and hazes reduce the amplitude of transmission spectra and thus decrease the molecular spectral features with a negative effect on the determination of molecular abundances. The emission spectra are less affected by the clouds and hazes \citep[e.g.,][]{Line16}. The spectral absorption depth of H$_2$S in the emission spectra is much greater than the expected noise level. Therefore, the H$_2$S feature can be detected using the emission spectra for atmospheres with clouds and haze. However, the absorption depth of PH$_3$ in the emission spectra is smaller than the expected noise level, therefore, the detection of PH$_3$ will be difficult if the atmosphere is covered by clouds or hazes. 

\item We neglect the effect of photochemistry on the primary and secondary transit spectra. Photochemistry affects the spectra in two ways. First, photochemistry changes the abundance profiles in the upper atmospheres. The effect on the secondary transit spectra is expected to be small since the absorption in the planetary emission spectra occurs near the 1 bar level. There may be some effects on the primary transit spectra since the light travels a longer path in the transmission spectra than in the emission spectra. Most absorption should still be from more abundant molecules (e.g. H$_2$O, CH$_4$, CO, NH$_3$, H$_2$S, PH$_3$) in the atmospheres. The photochemical products (C$_2$H$_6$, C$_2$H$_2$, HCN) may contribute a small amount of absorption and it is unclear whether JWST is able to detect these photochemical species. The second effect of photochemistry is the production of hazes. The flat transmission spectra for hot Jupiters and super-Earth may be caused by the photochemical hazes in the upper atmospheres. The effect of hazes on the spectra is the 
shrinking of the spectral amplitude, making the detection of molecules more difficult.

\item We assumed a single value of eddy diffusion coefficient in all our simulations. The strength of vertical mixing is highly uncertain on exoplanets. In Fig. \ref{fig:comparison_K}, we compare the abundances calculated using $K_{\rm eddy}$ = $1 \times 10^9$ cm$^2$ s$^{-1}$ and $1 \times 10^8$ cm$^2$ s$^{-1}$. Molecular species such as H$_2$O and CH$_4$ are largely unaffected by the strength of vertical mixing since they are the dominant species throughout the atmosphere. For disequilibrium species such as CO and CO$_2$, the effect is to shift the abundances in the same direction. CO is increased by less than one order of magnitude due to the change of $K_{\rm eddy}$. 

\item In this paper, we restrict our study to solar composition atmospheres. However, the elemental composition of exoplanetary atmospheres can be diverse. Jupiter's atmosphere is enriched in heavy elements relative to solar. It is reasonable to assume extrasolar giant planets have similar enrichment. If all the  heavy elements (C,N,O,S,P) are enriched similarly, the shape of the abundance profiles is preserved with only an upward shift. We expect the transmission spectra to have smaller spectral amplitudes since the pressure scale height is expected to be smaller for higher molecular mass atmospheres. This has an adverse effect on detecting molecules. However, higher mean molecular weight often correlates with smaller mass. For Neptune-size planets, the smaller gravity means higher scale height, and larger spectral amplitudes. The opposite effect of gravity and molecular mass on the spectra should rely on detailed modeling of Neptune -size exoplanets, which will be discussed in our next paper. 
If carbon and oxygen are not similarly enriched, for example C/O different than solar, the composition will be dramatically different for hot atmospheres. 

\end{itemize}

\section{Conclusions}\label{sec:conclusions}

In this paper, we modeled phosphorus and sulfur chemistry in the solar composition atmospheres of Jupiter-size planets with different levels of insolation. 
We find PH$_3$ is the primary carrier of phosphorus for atmospheres with $T_{\rm eq}$ $<$ 1000 K; P$_2$ is the primary carrier of phosphorus for $T_{\rm eq}$ greater than 1000 K and smaller than 1500 K; PH and PH$_2$ are the primary phosphorus bearing species for $T_{\rm eq}$ $>$ 2000 K. H$_2$S is the primary carrier of sulfur for $T_{\rm eq}$ $<$ 2000 K. We also compute the abundance profiles of major H/C/N/O/S bearing species. With the computed vertical profiles for H$_2$O, CO, CO$_2$, CH$_4$, NH$_3$, N$_2$, HCN, H$_2$S, and PH$_3$, we compute the synthetic primary and secondary transit spectra. 
We focus on identifying the spectral features for H$_2$S, PH$_3$, and HCN. We find spectral features of PH$_3$ at 4.0 $\sim$ 4.8 $\micron$,  H$_2$S at 2.5 $\sim$ 2.8 $\micron$ and 3.5 $\sim$ 4.1 $\micron$, HCN at 12 $\sim$ 16 $\micron$. We then simulated the noise for transits with K = 6.8 G star, and evaluated the detectability of PH$_3$ and H$_2$S with JWST observations. We find PH$_3$ can be detected for solar composition atmospheres with $T_{\rm eq}$ $\sim$ 500 K using the NIRCam LW F444W mode with a total observing time of 28.8 hrs. We find H$_2$S can be detected for solar composition atmospheres with $T_{\rm eq}$ $>$ 1500 K using the NIRCam LW grism F322W2 mode with a total observing time of 24.0 hrs. The detection of PH$_3$ and H$_2$S is complicated by the presence of clouds and hazes. In this case, H$_2$S may still be detected in the emission spectra, but PH$_3$ is difficult to detect due to the diminished spectra features by clouds and hazes.  

\acknowledgments

D.W. and J.I.L. acknowledge support from the Juno and JWST projects. Y.M. greatly appreciates the CNES post-doctoral fellowship program.

%

\software{ SAO98 \citep{ts76,tj02,kt09}, PandExo \citep{Batalha17} }

\bibliographystyle{aasjournal}
\bibliography{master}




\begin{deluxetable}{ccccccccccc}
\tablecaption{Parameters for computing noise - extension of Table 4 in \citet{Greene16} \label{tab:noise_params}}
\tablehead{
\colhead{Instrument} & 
\colhead{Mode} &  
\colhead{$\lambda$ ($\micron$)} & 
\colhead{native $R$} &
\colhead{$A_{\rm pix}$ (arcsec$^2$)\tablenotemark{a}} & 
\colhead{$n_{\rm pix}$\tablenotemark{b}} & 
\colhead{$b$ (e$^{-}$ s$^{-1}$ arcsec$^{-2}$)\tablenotemark{c}} & 
\colhead{$N_d$ (CDS)} & 
\colhead{$n_{\rm groups}$} &
\colhead{noise floor (ppm)\tablenotemark{d}} 
}
\startdata
\rotate
NIRCam & LW grism F322W2 & 2.5$-$3.9 & 1200@2.5 $\micron$ & 0.064"$\times$0.064" 
& 4 & 203.62 & 18 & 12 & 30\\
NIRCam & LW grism F444W & 3.9$-$5.0 & 1550@4.5 $\micron$ & 0.064"$\times$0.064"
& 4 & 308.74 & 18 & 25 & 30\\
MIRI & Slitless LRS prism & 5.0$-$11.0 & 40@5$\micron$, 160@10$\micron$ &  0.110"$\times$0.110"
& 4 & 5156.0 & 46 & 8 & 50\\
\enddata
\tablenotetext{a}{\url{http://www.stsci.edu/jwst/instruments/}}
\tablenotetext{b}{two times the spatial extent of point source spectrum}
\tablenotetext{c}{The background photon rate is computed 
at the JWST exposure time calculator \url{https://demo-jwst.etc.stsci.edu/}
}
\tablenotetext{d}{The adopted noise floor values are from \citet{Greene16}.}
\end{deluxetable}

\clearpage
\begin{deluxetable}{cccccccccc}
\tablecaption{Fiducial planetary system parameters in the model \label{tab:system_params}}
\tablehead{
\colhead{$T_{eq}$ (K)} & \colhead{$M_p$ ($M_{J}$)} & 
\colhead{$R_p$ ($R_J$)} & \colhead{$a_{\rm semi}$ (AU)} &
\colhead{$P$ (days)} &
\colhead{$T_{\star}$ (K)} & \colhead{$R_{\star}$ ($R_{\odot}$)} & \colhead{$T_{14}$ (s)} & \colhead{$D$ (pc)} & 
\colhead{$K$ (mag)} 
}
\startdata
500 & 1.0 & 1.0 & 0.310 & 63.1 & 5700 & 1.0 & 26000 & 50 & 6.8\\
750 & 1.0 & 1.0 & 0.138 & 18.7 & 5700 & 1.0 & 17333 & 50 & 6.8\\
1000 & 1.0 & 1.0 & 0.0776 & 7.89 & 5700 & 1.0 & 13000 & 50 & 6.8\\
1500 & 1.0 & 1.0 & 0.0345 & 2.34 & 5700 & 1.0 & 8666 & 50 & 6.8\\
2000 & 1.0 & 1.0 & 0.0194 & 0.986 & 5700 & 1.0 & 6500 & 50 & 6.8\\
\enddata
\end{deluxetable}

\clearpage 

\begin{deluxetable}{cccccc}
\tablecaption{Observing parameters for the planetary system in Table \ref{tab:system_params} with $T_{\rm eff}$ = 500 K. \label{tab:observing_500K}}
\tablehead{
\colhead{} & 
\colhead{ No. of groups per integration} & 
\colhead{ No. of integrations per occultation} & 
\colhead{ No. of transits} &
\colhead{observing efficiency} &
\colhead{transit duration} 
}
\startdata
NIRCam F322W2 & 12 & 11744 & 2 & 84.6\% & 7.2 hrs \\
NIRCam F444W & 25 & 5872 & 2 & 92.3\% & 7.2 hrs \\
MIRI LRS  & 8 & 36340 & 2 & 77.8\% & 7.2 hrs \\
\enddata
\end{deluxetable}

\begin{deluxetable}{cccccc}
\tablecaption{Observing parameters for the planetary system in Table \ref{tab:system_params} with $T_{\rm eff}$ = 1500 K. \label{tab:observing_1500K}}
\tablehead{
\colhead{} & 
\colhead{ No. of groups per integration} & 
\colhead{ No. of integrations per occultation} & 
\colhead{ No. of transits} &
\colhead{observing efficiency} &
\colhead{transit duration} 
}
\startdata
NIRCam F322W2 & 12 & 3916 & 5 & 84.6\% & 2.4 hrs \\
NIRCam F444W & 25 & 1958 & 5 & 92.3\% & 2.4 hrs \\
MIRI LRS  & 8 & 12112 & 5 & 77.8\% & 2.4 hrs \\
\enddata
\end{deluxetable}
\clearpage


\clearpage

\end{document}